\newif\iflongversion
\longversiontrue

\iflongversion
\documentclass[sigconf,nonacm,a4paper]{acmart}
\else
\documentclass[sigconf]{acmart}
\fi

\copyrightyear{2025}
\acmYear{2025}
\setcopyright{cc}
\setcctype{by-nc-sa}
\acmConference[CCS '25]{Proceedings of the 2025 ACM SIGSAC Conference on Computer and Communications Security}{October 13--17, 2025}{Taipei, Taiwan}
\acmBooktitle{Proceedings of the 2025 ACM SIGSAC Conference on Computer and Communications Security (CCS '25), October 13--17, 2025, Taipei, Taiwan}
\acmDOI{10.1145/3719027.3765128}
\acmISBN{979-8-4007-1525-9/2025/10}

\usepackage{graphicx} 
\usepackage[T1]{fontenc}
\usepackage[utf8]{inputenc}
\usepackage[inline]{enumitem}
\usepackage{listings}
\usepackage{booktabs}
\usepackage{tabularx}
\usepackage{multirow}
\usepackage{amsmath}
\usepackage{siunitx}
\usepackage{ragged2e}

\acmBadgeR[https://www.acm.org/publications/policies/artifact-review-and-badging-current]{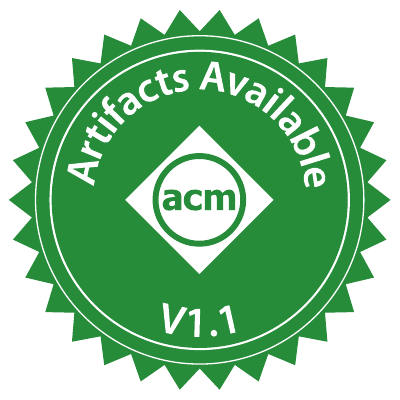}
\acmBadgeR[https://www.acm.org/publications/policies/artifact-review-and-badging-current]{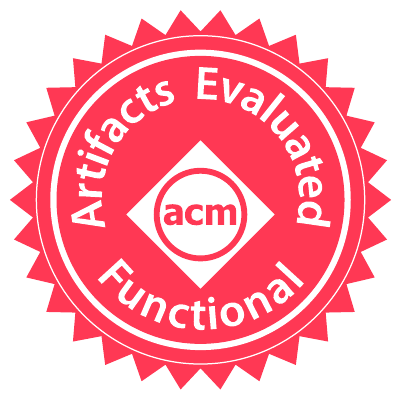}
\acmBadgeR[https://www.acm.org/publications/policies/artifact-review-and-badging-current]{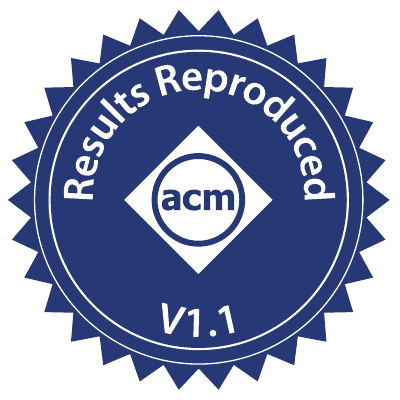}

\lstdefinestyle{actionstyle}{
    basicstyle=\ttfamily\footnotesize,
    breakatwhitespace=false,
    captionpos=b,
    keepspaces=true,
    showspaces=false,
    showstringspaces=false,
}

\usepackage{tikz}
\DeclareRobustCommand*\circled[1]{\tikz{\node[shape=circle,fill=black,minimum height=0.75em,text height=0.52em,inner sep=0.1pt] (char) {\textcolor{white}{\textbf{#1}}}}}

\iflongversion 
\fi 

\begin{document}

\title{Attestable Builds: Compiling Verifiable Binaries on Untrusted Systems using Trusted Execution Environments}

\author{Daniel Hugenroth}
\authornote{These authors contributed equally to this work.}
\email{dh623@cst.cam.ac.uk}
\orcid{0000-0003-3413-1722}
\affiliation{%
  \institution{University of Cambridge}
  \city{Cambridge}
  \country{United Kingdom}
}

\author{Mario Lins}
\authornotemark[1]
\email{mario.lins@ins.jku.at}
\orcid{0000-0003-1713-3347}
\affiliation{%
  \institution{Johannes Kepler University Linz}
  \city{Linz}
  \country{Austria}
}

\author{René Mayrhofer}
\email{rm@ins.jku.at}
\orcid{0000-0003-1566-4646}
\affiliation{%
  \institution{Johannes Kepler University Linz}
  \city{Linz}
  \country{Austria}
}

\author{Alastair R. Beresford}
\email{arb33@cst.cam.ac.uk}
\orcid{0000-0003-0818-6535}
\affiliation{%
  \institution{University of Cambridge}
  \city{Cambridge}
  \country{United Kingdom}
}


\begin{abstract}
In this paper we present attestable builds, a new paradigm to provide strong source-to-binary correspondence in software artifacts.
We tackle the challenge of opaque build pipelines that disconnect the trust between source code, which can be understood and audited, and the final binary artifact which is difficult to inspect.
Our system uses modern trusted execution environments (TEEs) and sandboxed build containers to provide strong guarantees that a given artifact was correctly built from a specific source code snapshot.
As such it complements existing approaches like reproducible builds which typically require time-intensive modifications to existing build configurations and dependencies, and require independent parties to continuously build and verify artifacts.
In comparison, an attestable build requires only minimal changes to an existing project, and offers nearly instantaneous verification of the correspondence between a given binary and the source code and build pipeline used to construct it.
We evaluate it by building open-source software libraries—focusing on projects which are important to the trust chain and have proven difficult to be built deterministically.
The overhead (42\,seconds start-up latency and 14\% increase in build duration) is small in comparison to the overall build time.
Importantly, our prototype can build complex projects such as LLVM Clang without requiring any modifications to their source code and build scripts.
Finally, we formally model and verify the attestable build design to demonstrate its security against well-resourced adversaries.
\end{abstract}


\maketitle

\section{Introduction}
\label{sec:introduction}
Executable binaries are digital black boxes.
Once compiled, it is hard to reason about their behavior and whether they are trustworthy.
In contrast, source code is easier to inspect.
However, few have the ability, resources, and patience to compile all their software from scratch.
Therefore, we seek to enable recipients to verify that an artifact has been truthfully built from a given source code snapshot.
This challenge has been popularized in the now-famous Turing Lecture by Ken Thompson on ``Trusting Trust''~\cite{thompson1984trustingtrust}.

The problem of trusting build artifacts also presents itself in commercial settings where source code is typically not shared.
In this context, the source code is the only source-of-truth that can be inspected by the employed engineers and auditors:
During code review it is code changes and not binary output that is examined and, likewise, audit reports generally reference repository commits and not the hash of the shipped artifact.
Hence, \textit{verifiable source-to-binary correspondence} is also relevant in an enterprise setting.
Where this correspondence cannot be verified, defects are difficult to identify---allowing them to spread down the supply chain to many targets.

In recent years, adversaries have successfully targeted build processes.
During the 2020 SolarWinds hack, attackers compromised the company's build server to inject additional code into updates for network management system software~\cite{wired2023solarwind}.
As there were no changes to the source code repository, only forensic inspection of the build machines eventually unveiled the malicious change.
In the meantime, the software was distributed to many customers in industry and government that relied on it to secure access to their internal networks.
The US Cybersecurity \& Infrastructure Security Agency (CISA) issued an emergency directive requesting immediate disconnect of all potentially affected products~\cite{cisa2020ed2101}.

In 2024 a complex supply chain attack (CVE-2024-3094) against the XZ Utils package was uncovered that allowed adversaries to compromise vulnerable servers running OpenSSH~\cite{lins2024libxz}.
A key enabler for this attack is that the maintainers of (open-source) projects utilizing Autoconf often manually create certain build assets (e.g., a configure script), add it to a tarball, and then provide it to the packager, who builds the final artifact. 
In case of XZ, this tarball contained a malicious asset covertly included by the adversary that was not part of the repository. 
Here both the maintainer and the packager have opportunity to meddle with the final binary artifact.

\textit{Reproducible Builds} (R-Bs, \S\ref{sec:background:reproducible-builds}) are the typically proposed solution to address potential discrepancies between source code and compiled binaries.
Correctly implemented, R-Bs ensure source-to-binary correspondence by making the build process perfectly deterministic.
Thus, they guarantee that the same source code always results in a bit-to-bit identical binary artifact output.
This enables independent parties to reproduce binary artifacts, thus verifying that a given source input generated a given output.
There are many successful projects that implement R-Bs~\cite{debianrb, torrb, chromiumrb}.

However, R-Bs come with their own challenges, requiring time-intensive and therefore costly changes to the build process. These are incurred not just as a one-time cost but as a continuous maintenance burden.
Further, for closed-source software, the downstream consumer cannot check if their supplier has correctly applied R-B principles, since they are typically not given access to the required source code.
Additionally, even for source-available software, the build process and compiler are often not available, for example due to intellectual property or licensing concerns.
In reality, R-Bs only provide effective security benefits when there are independent builders who are continuously verifying that distributed artifacts are identical to their locally built ones.

We propose \textit{Attestable Builds} (A-Bs) as a practical and scalable alternative where R-Bs are infeasible or costly to implement---including as a complement to extend R-B guarantees to consumers who cannot verify R-Bs themselves even if the primary build chain has R-B properties.
For this we leverage Trusted Execution Environments (TEEs) to ensure that the build process is performed correctly and is verifiable.
Unlike previous generations of TEEs (e.g., Intel SGX, Arm TrustZone), modern TEE implementations (e.g., AMD SEV-SNP, Intel TDX, AWS Nitro Enclaves) support full virtual machines with strong protection against interference by the hypervisor and physical attacks.
Whereas this technology is typically used to achieve data confidentiality, in this work we leverage its integrity properties.
This has an additional security benefit: Since integrity attacks are inherently active attacks, this limits the window of opportunity for attacks.

A-Bs are compatible with the reality of modern software engineering practices and allow the build process to be performed by an untrusted build service run by an untrusted cloud service provider (CSP), as long as the TEE hardware is trusted.
The idea of A-Bs also extends to other computed artifacts beyond compiled binaries: For instance, A-Bs can additionally attest to test results and static analyses for additional guarantees about the artifacts (\S\ref{sec:discussion}).
Table~\ref{tab:rb-versus-ab} highlights the similarities and differences between R-Bs and A-Bs.
We believe that A-Bs would have prevented or substantially mitigated the feasibility of the mentioned SolarWind and XZ Utils attacks and/or helped to detect them more easily (\S\ref{sec:discussion:case-studies}).

The overall A-B design is simple:
First, an open-source machine image boots inside a modern TEE.
The TEE guest then downloads the source code repository and commits to a hash of the downloaded files, including build instructions, in a secure manner before executing the build process inside a sandbox.
Afterwards, the TEE hardware trust anchor attests to the booted image, the committed hash value, and the build artifact.
The resulting attestation certificate is shared alongside the artifact and is recorded in a transparency log.
Finally, the recipient of the artifact can inspect the certificate locally and fetch the corresponding entry from the transparency log to verify that a given artifact has been built from a particular source code snapshot.

\begin{table}[t]
    \centering
    \renewcommand{\arraystretch}{1.3}
    \caption{Comparison of Reproducible Builds (R-Bs) and Attestable Builds (A-Bs).}
    
    \begin{tabular}{p{0.455\linewidth}|p{0.455\linewidth}}
        \toprule
         \multicolumn{1}{c|}{\textbf{Reproducible Builds}} & \multicolumn{1}{c}{\textbf{Attestable Builds}} \\
         \midrule 
         \multicolumn{2}{c}{\circled{+}\;Strong source-to-binary correspondence}\\
         \midrule
         \circled{-}\;High engineering effort for both initial setup and ongoing build maintenance & \multirow{ 2}{*}{\shortstack[l]{\circled{+}\;Only small changes to the\\build environment needed\\[4pt]\circled{+}\;Cloud service compatible}}\\  
         \circled{-}\;Dependencies and tool chain need to be deterministic & \circled{~}\;Dependencies and tool chain can be R-B or A-B \\
         \circled{-}\;Environment might leak into build process undetected & \circled{+}\;Enforces hermetic builds\\
         \circled{+}\;Machine independent & \circled{-}\;Requires modern CPU\\
         \circled{~}\;Requires trusting at least one party and their machine & \circled{~}\;Requires trusting the hardware vendor\\
         \circled{-}\;Requires open source & \circled{+}\;Supports closed source and signed intermediate artifacts\\
         \midrule 
         \multicolumn{2}{c}{ \circled{+}\;Can be composed to an anytrust setup~(\S\ref{sec:stamp:composing})}\\
         \bottomrule
    \end{tabular}
    \label{tab:rb-versus-ab}
\end{table}

One important insight of our work is that nested sandboxing is required because of the current shortcoming of hardware based enclaves such as AMD SEV-SNP (or Amazon Nitro).
For these, the remote attestation guarantees stop at boot time and they do not provide nested enclaves.
That is, while the hardware security primitives guarantee to protect the host environment from interference by the guest VM---and vice versa---in terms of memory confidentiality and integrity, there is no guarantee on the run-time state within each VM enclave.
However, A-Bs necessarily process and execute untrusted and potentially malicious code within a running VM.
Therefore, nested sandboxing is required to allow treating the build process as an untrusted black box within the VM, which could otherwise compromise integrity assumptions of the output artifacts.
These properties cannot currently be achieved by either hardware VM enclaves or containerization alone---only in combination.
As such, our work motivates the need for nested enclave support which provide much stronger integrity guarantees than an inner-nested sandbox.

We also present a composition of A-Bs and R-Bs that achieves novel trust properties (\S\ref{sec:stamp:composing}).
By executing R-Bs inside enclaves from different TEE vendors, consumers can trust the artifact as long as they trust any of the vendors---without having to explicitly decide which one they trust, i.e., an any-trust model.
In a typical R-B setup, users cannot easily verify the particular build environment of the involved parties and whether they might share common vulnerabilities, e.g. the same backdoored CPU firmware.
As such, this work also contributes to a better understanding of R-B setups and presents an approach to strictly improve their guarantees.

Finally, we provide a specific threat model of cloud-based build services in relation to A-Bs and R-Bs (\S\ref{sec:stamp:threatmodel}) highlighting the underlying trust assumptions, adversary models, and threats.
Based on this, we then formally model and verify our protocol using Tamarin~\cite{tamarin_2025}, a security protocol verification tool, to show that A-Bs provide relevant security guarantees (\S\ref{sec:formalverification}).

In this paper, we make the following contributions:
\begin{itemize}
    \item We present a new paradigm called Attestable Builds (A-Bs) that provides strong source-to-binary correspondence with transparency and accountability. We discuss the shortcomings of alternative approaches and devise a design relying on a sandbox and an integrity-protected observer.
    \item We describe a novel composition of A-Bs and R-Bs which provides trustworthy provenance for compiled artifacts in a strong any-trust model.
    \item We implement an open-source prototype to demonstrate the practicality of A-Bs by building real-world software including complex projects like Clang and the Linux Kernel as well as packages that are hard to build reproducibly.
    \item We evaluate the performance of our system and find that it adds a (mitigable) 42\,second start-up cost, which is small compared to typical build durations. It also imposes a performance overhead of around 14\% in our default configuration and up to 68\% when using hardened sandboxes.
    \item We provide threat modeling for verifiable build paradigms such as A-Bs based on Confidential Computing and discuss how it can be extended to other tasks such as compliance tests and static analysis.
    \item We formally verify the system using Tamarin and discuss the underlying trust assumptions required.
\end{itemize}

\section{Background}
\label{sec:background}
Attestable builds integrates with modern software engineering and CI/CD patterns (\S\ref{sec:background:swe-cicd}) and provides an alternative to R-Bs (\S\ref{sec:background:reproducible-builds}).
For this we leverage Confidential Computing technology (\S\ref{sec:background:confidential-computing}) and verifiable logs (\S\ref{sec:background:verifiable-logs}).
This section introduces the required background and building blocks.

\subsection{Modern software engineering \& CI/CD}
\label{sec:background:swe-cicd}

Modern Software Engineering (SWE) involves large teams that requires efficient mediation of their collaboration aspects through software.
Many projects rely on source control management (SCM) software like Git~\cite{git} and Mercurial~\cite{mercurial}.
The underlying repositories are often hosted by online services, such as GitHub~\cite{github} or Bitbucket~\cite{bitbucket}.
We call these Repository Hosting Providers (RHPs).

With increasing complexity, Continuous Integration (CI), has become an important component in modern software projects.
Every published code change triggers a new execution of the project's CI pipeline that builds, tests, and verifies the new code snapshot.
In addition, some code changes might trigger a (separate) Continuous Deployment (CD) pipeline which after passing all checks distributes binaries automatically and re-deploys them to the production system.
Such CI/CD pipelines are described in configuration files within the source code repository and then executed by online services, build service providers (BSPs), such as Jenkins~\cite{jenkins} or GitHub Actions~\cite{githubactions}.
The latter is an example where the RHP is also a BSP.
Our prototype uses GitHub Actions to demonstrate how A-Bs can integrate into existing infrastructure (\S\ref{sec:eval:implementation}).

\begin{figure}
    \centering
    \includegraphics[width=0.8\columnwidth]{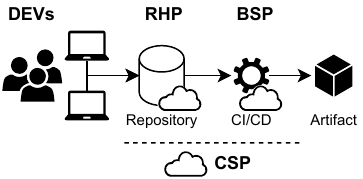}
    \caption{Developers (DEVs) commit to a source code repository at a repository hosting provider (RHP). Changes trigger the CI/CD pipeline at a build service provider (BSP) and generate new binary artifacts. RHP and BSP typically run on servers provided by a cloud service provider (CSP).}
    \label{fig:swe-model}
\end{figure}

Both RHPs and BSPs often do not manage their own machines, but use cloud infrastructure provided by cloud service providers (CSPs) such as Amazon Web Services (AWS), Microsoft Azure, or Google Cloud Platform (GCP).
Although there are self-hosted alternatives, such as GitLab~\cite{gitlab}, even those are often deployed via a CSP.
Figure~\ref{fig:swe-model} shows the involved parties.

\subsection{Reproducible builds (R-Bs)}
\label{sec:background:reproducible-builds}

The use of CI/CD brings many benefits to developers:
automated checks ensure that no ``broken code'' is checked in,
builds are easily repeatable since they are fully described in versioned configuration files,
and long compile/deploy cycles happen asynchronously.
However, they also shift a lot of trust to the RHP, BSP, and CSP.
These online services are opaque and any of them can interfere with the build process.
Therefore, the conveniently outsourced CI/CD pipeline undermines the trustworthiness of the generated artifacts.
This leads to a particularly undesirable situation, as its binary output is hard to inspect and understand.
Therefore, trust in the process itself is just as important as trust in its input.

R-Bs have been proposed as a solution to ensure source-to-binary correspondence.
The underlying approach is to make the build process fully deterministic such that the same source input always yields perfectly identical binary output.
In a project with R-Bs, malicious build servers can be uncovered by repeating the build process on a different machine.
Correctly set up, the builds are replicated by independent parties that then compare their results.

However, introducing R-Bs to a software project is challenging~\cite{fourne2023cispa,shi2021experience,butler2023business}.
For bit-to-bit identical outputs, the build process needs to be fully described in the committed files and all steps need to be fully deterministic.
However, sources of non-determinism are plentiful as outputs can be affected by timestamps, archive metadata, unspecified filesystem order, build location paths, and uninitialized memory~\cite{fourne2023cispa,shi2021experience}.

While many sources of non-determinism can be eliminated with effort and tooling, other steps, such as digital signatures used to sign intermediate artifacts in multi-layered images, cannot easily be made deterministic.
This is because typical signature algorithms break when random/nonce parameters become predictable and might leak private key material as a result~\cite{katz2007introduction}.
For example, consider a build process for a smartphone firmware image that builds a signed boot loader during its process.
This inner signature will affect the following build artifacts and is not easily hoisted to a later stage.
In other instances, this signing process might happen by an external service or in a hardware security module (HSM) to protect the private key and therefore can never be deterministic.

Critically, for the downstream package to be reproducible, all its dependencies need to be reproducible as well.
This also applies for dependencies that are shipped as source code, as R-B is a property of the build system.
Facing non-determinism in any of the (transitive) upstream dependencies, a developer either needs to fix the upstream dependency or fork the respective sub-tree.
In practice, the verification of having achieved R-B is often done heuristically and newly identified sources of non-determinism can cause a project to loose its status~\cite{fourne2023cispa}.
Despite the challenges, there are large real-world projects that have successfully adopted R-Bs. 
Examples are Debian~\cite{debianrb}, NetBSD~\cite{netbsdrb}, Chromium~\cite{chromiumrb}, and Tor~\cite{torrb}.
However, these came at considerable expenses in terms of required upgrades to the build system and on-going maintenance costs~\cite{lamb2021reproduciblefoss,fourne2023cispa}.

The Debian R-B project stands out due to its scale and highlights the challenges of R-Bs, taking twelve years to produce the first fully reproducible Debian image~\cite{debian_reproducibility,lwn2025debianrepoiso}.
A typical challenge is to motivate upstream developers to provide reproducible packages.
This led to the introduction of a bounty system offering prioritized inclusion if a build is reproducible~\cite{debian_reproducibility}.
The project's dashboard~\cite{debian_developer_dashboard} shows that the number of unreproducible packages dropped from 6.1\% (Stretch, released 2017) to 2.0\% (Bookworm, released 2023).
This suggests that the remaining packages are particularly difficult to convert to R-Bs.
Therefore, we picked some of these packages for our practical evaluation (\S\ref{sec:eval:implementation}).

\subsection{Confidential Computing}
\label{sec:background:confidential-computing}

Executing code in a trustworthy manner on untrusted machines is a long standing challenge.
Enterprises face this challenge when processing sensitive data in the cloud and financial institutions need to establish trust in installed banking apps.
These scenarios require a solution that ensures that the data is not only protected while in-transit or at-rest, but also when in-use.
Trusted Execution Environments (TEEs) allow the execution of code inside an \emph{enclave}, a specially privileged mode such that execution and memory are shielded from the operating system and hypervisor.
Typically, the allocated memory is encrypted with a non-extractable key such that it resists even a physical attack with probes used to intercept communication between CPU and RAM (and potentially interfere with).
Even the hypervisor can only communicate with the enclaves via dedicated channels, e.g., \textit{vsock} or shared memory.
However, the hypervisor maintains the ability to pause or stop code execution inside an enclave.

Earlier technologies such as ARM~TrustZone~\cite{pinto2019armtrustzoneeplained} and Intel~SGX~\cite{costan2016intelsgxexplained} create enclaves on a process level.
This requires application developers to rewrite parts of their application using special SDKs so secure functionalities are run inside an enclave.
In particular, Intel~SGX has proven to be vulnerable to side-channel attacks that allow adversaries to extract secret information from enclaves~\cite{chen2019sgxpectre,skarlatos2019microscope,murdock2020plundervolt,van2020sgaxe}.
It also imposes further practical limitations, such as a maximum enclave memory size and performance overhead.

More recent technologies such as Intel~TDX~\cite{inteltdx} and AMD~SEV-SNP~\cite{amdsevsnp} boot entire virtual machines (VMs) in a confidential context.
This promises to simplify the development of new use-cases as existing applications and libraries can be used with little to no modification.
In addition, VMs can be pinned to specified CPU cores, reducing the risk of timing and cache side-channel attacks.
AWS Nitro is a similar technology, built on the proprietary AWS  Nitro hypervisor and dedicated hardware.
The trust model is slightly weaker as the trusted components sit outside the main processor.
We choose AWS Nitro for our prototype due to its accessible tooling, but it can be substituted with equivalent technologies.

It is important for the critical software to verify that it is  running inside a secure enclave.
Likewise, users and other services interacting with critical software need to verify the software is running securely and is protected from outside interference and inspection.
This is typically achieved using \textit{remote attestation}.
On a high-level, the curious client presents a challenge to the software that claims to run inside an enclave.
The software then forwards this challenge to the TEE and its backing hardware who signs the challenge and binds it to the enclave's Platform Configuration Registers (PCRs).
The PCRs are digests of hash-chained measurements that cover the boot process and system configuration that claims to have been started inside the TEE~\cite{tpmstandard}.

It is typically not possible to run an enclave inside another enclave or to compose these in a hierarchical manner---although new designs are being discussed~\cite{ghosn2023creating}.
This presents a challenge in our case as we need to run untrusted code, i.e. the build scripts stored in the repository, inside the enclave.
We work around this technical limitation by sandboxing those processes inside the TEE.

\subsection{Verifiable logs}
\label{sec:background:verifiable-logs}

A verifiable log~\cite{eijdenberg2015verifiabledatastructure} incorporates an append-only data structure which prevents retroactive insertions, modifications, and deletions of its records.
In summary, it is based on a binary Merkle tree and provides two cryptographic proofs required for verification.
The inclusion proof allows verification of the existence of a particular leaf in the Merkle tree, while the consistency proof secures the append-only property of the tree and can be used to detect whether an attacker has retroactively modified an already logged entry.
While such a transparency log is not strictly necessary to verify the attested certificate of an artifact, it adds additional benefits such as ensuring the distribution of revocation notices, e.g., after discovering vulnerabilities or leaked secrets.
Artifacts providers can also monitor it to detect when modified versions are shared or their signing key is being used unexpectedly.
A central log can also be used to include additional information, such as linking a security audit to a given source code commit (\S\ref{sec:discussion}).

\section{Attestable Builds (A-Bs)}
\label{sec:stamp}
This section introduces the involved stakeholders, the considered threat model, the design of a typical A-B architecture, and how it can be composed with R-Bs.

\subsection{Stakeholders}
\textbf{The verifier} receives an artifact, e.g., an executable, either directly from a specific BSP or via third-party channels.
This could be a user downloading software or a developer receiving a pre-built dependency from a package repository.
In general, the verifier does not trust the CI/CD pipeline and therefore wants to verify the authenticity of the respective artifact.
\textbf{The artifact author}, e.g., a developer or a company, regularly builds artifacts for their project and distributes them to downstream participants.
Thus, the system should integrate with existing version control systems hosted by an RHP.
The artifact author also does not trust the CI/CD pipeline, as they do not control the involved hardware.
Therefore, they need to detect any unauthorized manipulation.
\textbf{All other stakeholders} (RHP, BSP, CSP, HSP, \dots) are untrusted.
We assume there are no restrictions on combining multiple roles on one stakeholder, which is the realistic and more difficult set-up as it makes interference less likely to be detected.
For example, a self-hosted GitLab operator would take over the role as RHP to manage the source code using git, the BSP by providing build workflows, and the CSP by providing the underlying servers that execute the build steps.
Only for the transparency log we require a threshold of honest operators, e.g., in the form of independent witnesses tracking the consistency of the log similar as it is already done in  established infrastructure such as Certificate Transparency~\cite{laurie2014certificate} and Sigstore~\cite{newman_2022}.

\subsection{Threat model}
\label{sec:stamp:threatmodel}
The main security objective is to provide an attested build process with strong source-to-binary correspondence guarantees. 
We do not consider confidentiality or availability as security objectives in A-Bs, assuming that the source code is not inherently confidential and that ensuring availability of relevant components in the build pipeline is the responsibility of the infrastructure provider. 
However, since the TEEs can also provide confidentiality, A-Bs can be adapted accordingly.
Our threat model focuses on the build process as illustrated in Figure~\ref{fig:swe-model}, describing pipelines where an artifact author publishes code to a repository, which is then built and deployed by the BSP.  

\subsubsection{Assumptions}
\label{sec:stamp:threatmodel:assumptions}
We make the following assumptions for our threat model: we assume that the enclave itself is trusted, including the hardware-backed attestation provided by the TEE.
We later expand on the intricacies of this statement (\S\ref{sec:stamp:threatmodel:confcomp}) alongside our attack scenarios (\S\ref{sec:stamp:architecture:tee-attacks-and-impact}).
We assume that the transparency log is trustworthy as potential tampering attempts are detectable.
We also assume that the transparency log is protected against split-view attacks by having sufficient witnesses in place. 

In this paper, we refer to all components---hardware, firmware, and software---involved directly and indirectly in producing the final artifact as build dependencies.
In particular, these dependencies include the TEE firmware, the compilation tool-chain, the image running inside the TEE, and software libraries referenced by the source code and build configuration.
Dependencies can consist of or rely on other dependencies recursively.
Therefore, in order to make strong provenance claims about the built artifact, all build dependencies must have verifiable provenance, e.g., through R-Bs or A-Bs, to mitigate~\ref{threat:backdoor-in-repo}.
Otherwise, e.g., a backdoored compilation tool-chain could invalidate the artifact trust assumption (\S\ref{sec:stamp:architecture:build-dependencies}).

A-Bs rely on benign build dependencies for producing secure artifacts.
A malicious build dependency might yield an insecure artifact or otherwise tamper with the build process within the sandbox.
However, all build dependencies contribute to either the enclave measurements (firmware, base image) or the source code snapshot (hashes in lockfiles for external dependencies, vendored-in dependencies, ...).
Therefore, if a build dependency is later found to be malicious, the affected artifacts can be identified in the transparency log and then revoked.

\subsubsection{Adversary modeling}

The following list defines relevant adversary models, including information about the respective attack surface, in accordance to the scope of our research.

\label{sec:stamp:threatmodel:attackermodel}
\begin{enumerate}[start=1,label={A\arabic*}]
    \item \label{attacker:physical} \textbf{Physical adversary:}
    Adversary with physical access to hardware, including storage, or the respective infrastructure.
    We assume that a physical adversary could also be an insider (\ref{attacker:insider}), as our threat model does not distinguish between attacks that require physical access, regardless of whether the attacker is external or internal.
    \item \label{attacker:opa} \textbf{On-path adversary (OPA):} 
    An on-path adversary has access to the network infrastructure (e.g., via a machine-in-the-middle [MitM] attack) and is capable of modifying code, the attestation data, or the artifact sent within that network. 
    \item \label{attacker:insider} \textbf{Insider adversary:} 
    An insider adversary can be a privileged employee working with access to the platform layer such as the hypervisor of the CSP running the VMs or the hosting environment of the BSP.
    This category of adversary includes malicious service providers.
    Physical attacks are covered through \ref{attacker:physical}.
\end{enumerate}

\subsubsection{Threats}
\label{sec:stamp:threatmodel:threats}

We introduce threats for generic build systems that we considered while designing A-Bs.
The following section on architecture explains how A-Bs effectively mitigates these.

\begin{enumerate}[start=1,label={T\arabic*}]
    \item \textbf{Compromise the build server:} \label{threat:compromise-build-process}
    An adversary (\ref{attacker:physical}, \ref{attacker:insider}) might compromise the build server infrastructure by modifying aspects of the build process, including source code, which could result in a malicious build artifact. 
    This threat addresses all kinds of unauthorized modifications during the build process, such as directly manipulating the source code, the respective build scripts (e.g., shell scripts triggering the build), or parts of the build machine itself, like the OS. 

    \item \textbf{Cross-tenant threats:} \label{threat:bad-build}
    Any adversary that uses shared infrastructure might use its privilege to temporarily or permanently compromise the host and thus affect subsequent or parallel builds.
    It also potentially renders any response from the service untrustworthy.
    This is particularly important for build processes as they generally allow developers to execute arbitrary code.
    
    \item \textbf{Implant a backdoor in code or assets:} \label{threat:backdoor-in-repo}
    An adversary (\ref{attacker:insider}) might implant a backdoor within the repository through intentionally incorrect code or within files that are committed as binary assets.
    For this to be successful the adversary might need to successfully execute social engineering attack to become co-maintainer on an open-source repository.
    An example of this is the compromise of XZ Utils~\cite{lins2024libxz} which we discuss in Section~\ref{sec:discussion}.
    Unlike \ref{threat:compromise-build-process}, implanting a backdoor in this manner does not directly compromise the build process itself, but rather is an orthogonal supply chain concern.  
    
    \item \textbf{Spoofing the repository:} \label{threat:repository-spoofing}
    An adversary might clone an open-source project, introduce malicious modifications, and attempt to make it appear as the original repository as shown in recent attacks~\cite{spoofedgithubrepos}.
    This is similar to typo-squatting of dependencies in package managers~\cite{neupane2023beyond,taylor2020defending}.
    A common mitigation of such threats is the use of digital signatures for signing the artifact.
    However, an insider adversary (\ref{attacker:insider}) might be able to exfiltrate such a key.
    
    \item \textbf{Compromise build assets during transmission:} \label{threat:opa}
    An adversary with network access (\ref{attacker:opa}) might compromise build assets (e.g., source code, dependencies, compilation tool-chain, configuration, \dots) transferred between the parties involved in the build process by intercepting the network traffic.
    We consider well-resourced adversaries that might issue valid SSL certificates or compromise the servers of any other party.
    This threat does also include side-loading potentially malicious libraries from external sources. 
    
    \item \textbf{Compromise the hardware layer:} \label{threat:compromise-hardware} An adversary with physical access (\ref{attacker:physical}) might perform classical physical attacks such as interrupting execution, intercepting access to the RAM, and running arbitrary code on the CPU cores that are not part of a secure enclave. 
    This aligns with the threat model of Confidential Computing technologies although they all vary slightly and they do have known vulnerabilities.
    
    \item \textbf{Undermine verification results:} \label{threat:break-verification}
    An adversary (\ref{attacker:physical}, \ref{attacker:opa}, \ref{attacker:insider}) can undermine verification results, e.g., authenticity or integrity checks, by manipulating verification data either directly in the infrastructure or while in transit.   
    Similarly, an adversary (\ref{attacker:opa}) might pursue a split-view attack in which some users receive different results for queries against logs.

\end{enumerate}

\subsubsection{Confidential Computing}
\label{sec:stamp:threatmodel:confcomp}

For A-Bs, we require the underlying TEE technology to provide unforgeable remote attestation covering hardware, firmware, and the image running inside the enclave.
The security of the attestation is the most critical TEE guarantee as it later allows identifying and disapproving artifacts built in retrospectively-insecure environments (\S\ref{sec:stamp:architecture:tee-attacks-and-impact}).
In addition, we require strong integrity properties, i.e. the measured enclave code is executed correctly and isolated from the host system.
We do not require confidentiality for A-B.
So, many attacks targeting confidentiality, including many side-channel attacks, do not impact the provenance guarantees of artifacts built with A-Bs (\S\ref{sec:relatedwork:attacks}).
However, if confidentiality of source code and build configuration is desired, this can be added as an optional security goal.

\subsection{Architecture}
\label{sec:stamp:architecture}

\begin{figure*}[p!]
    \centering
    \includegraphics[width=\textwidth]{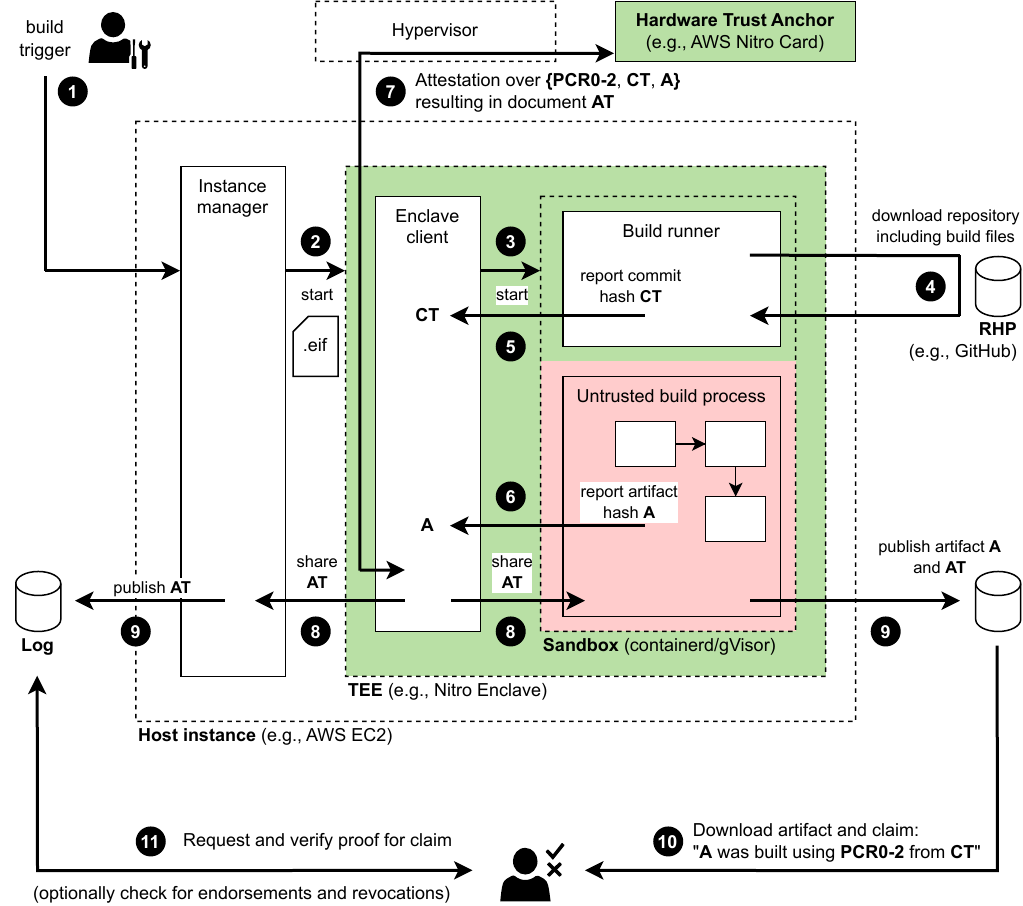}    
    \caption{Overview of the protocol steps during build and verification. Dashed borders indicate separate or sandboxed execution environment. Only the TEE and the hardware trust anchor are fully trusted.\\~\\
    $\vcenter{\hbox{\includegraphics[height=3.5mm]{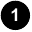}}}$ The build process is triggered manually or as a result of code changes. Either will cause a webhook call to the Instance Manager.
    $\vcenter{\hbox{\includegraphics[height=3.5mm]{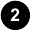}}}$ The Instance Manager starts an fresh enclave from a publicly known \texttt{.eif} file with the measurements PCR0-2.
    $\vcenter{\hbox{\includegraphics[height=3.5mm]{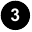}}}$ Once booted, the Enclave Client starts the inner sandbox.
    $\vcenter{\hbox{\includegraphics[height=3.5mm]{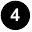}}}$ The sandbox executes the action runner which fetches the repository snapshot. That snapshot includes both the source code and build instructions.
    $\vcenter{\hbox{\includegraphics[height=3.5mm]{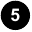}}}$ A hash of the snapshot is reported to the Enclave Client for safeguarding.
    Now the build process is started which is untrusted.
    $\vcenter{\hbox{\includegraphics[height=3.5mm]{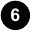}}}$ Once it finishes, the sandbox reports the hash of the produced artifact.
    $\vcenter{\hbox{\includegraphics[height=3.5mm]{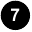}}}$ The Enclave Client then requests an attestation document from the Nitro Card covering PCR0-2, the repository snapshot hash, and the artifact hash.
    $\vcenter{\hbox{\includegraphics[height=3.5mm]{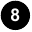}}}$ The results are shared with both the build process and the outer Instance Manager.
    $\vcenter{\hbox{\includegraphics[height=3.5mm]{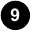}}}$ The build process can now publish the artifact and certificate. And the Instance Manager publishes the attestation.
    $\vcenter{\hbox{\includegraphics[height=3.5mm]{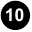}}}$ When a user downloads the artifact, it can contain a certificate specifying how it was build.
    $\vcenter{\hbox{\includegraphics[height=3.5mm]{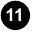}}}$ The user can verify this certificate by checking that it is included in the public transparency log.
    }
    \label{fig:protocol-steps}
\end{figure*}

We designed A-Bs with cloud-based CI/CD pipelines in mind.
In particular, such a system can be provided by a BSP who rents infrastructure from an untrusted CSP (see Figure~\ref{fig:swe-model}).
Our design is compatible with different Confidential Computing technologies.
While our practical implementation (\S\ref{sec:practicalevaluation}) uses a particular technology, we describe our architecture and its design challenges in general terms (e.g., TEE, sandbox).
Figure~\ref{fig:protocol-steps} provides an architectural overview which is described in more detail in this section.

The core unit of an A-B system is the host instance which runs control software, the instance manager, and can start our TEE.
Each build request is forwarded to an instance manager which then starts a fresh enclave from a public image inside the TEE.
These images are available as open source and therefore have known PCR values that can later be attested to.

The TEE provides both confidentiality and integrity of data-in-use through hardware-backed encryption of memory which protects it from being read or modified---even from adversaries with physical access, the host, and the hypervisor.
The enclave uses remote attestation to prove that it has booted a particular secure image in a secure context.
These guarantees mitigate \ref{threat:compromise-build-process} and are essential to the integrity of the final attestation.
However, it alone is not sufficient, as the build process might manipulate its internal state, and thus the state we are later attesting to.
Therefore, we introduce an integrity-protected observer, the \textit{Enclave Client}, that interacts with a sandbox embedded within the TEE.

Once the enclave has booted, it starts the Enclave Client.
As it runs inside the TEE, we can assume that it is integrity-protected.
The Enclave Client first establishes a bi-directional communication channel with the Instance Manager outside the TEE via shared memory.
Through this channel, the Instance Manager provides short-lived authentication tokens for accessing the repository at the RHP and receives updates about the build process.

The Enclave Client then manages a \textit{sandbox} inside the enclave.
The sandbox ensures that the untrusted build process (which might execute arbitrary build steps and code) cannot modify the important state kept by the Enclave Client.
In particular, we need to protect the initial measurement of the received source code files and build instructions.
This mitigates \ref{threat:bad-build}.
The sandbox optionally captures complete, attested, logs of all incoming and outgoing communication of the build execution, which can help audits and investigations.

Once the sandbox has started, the Enclave Client forwards a short-lived authentication token to the build runner inside the sandbox.
The build runner uses the token to fetch both the code and build instructions from the RHP.
Since the enclave has no direct internet access, all TCP/IP communication is tunneled via shared memory as well.
Upon downloading the source code and instructions, the sandbox computes the commit hash \textbf{CT} and reports to the Enclave Client.
The commit hash not only covers the content of the code and build instructions, but also the repository metadata.
This includes the individual commit messages which can include signatures with the developers private keys~\cite{gitsignatures}.
By checking and verifying these during the build steps, the system also attests to the origin of the source code, i.e. the latest developer implicitly signs-off on the current repository state at this commit.
This mitigates \ref{threat:opa}.

Once the commit hash has been committed to the Enclave Client, the sandbox starts the build process by executing the build instructions from the repository---and \textit{from that moment we consider the inner state sandbox untrusted}.
The sandbox expects the build process to eventually report the path of the artifact that it intends to publish.
Once the build process is complete, the sandbox computes the hash \textbf{A} of the artifact and forwards it to the Enclave Client.
Note: while the inner state of the sandbox is untrusted, the Enclave Client as an integrity-protected observer has safeguarded the input measurements (CT) from manipulation.
A ratcheting mechanism ensures that it will only accept CT once at the beginning from the build runner before any untrusted processes are started inside the sandbox.
The hash of the artifact (A) can be received from the untrusted build process as it will be later compared by the user against the received artifact.

The Enclave Client then uses the TEE attestation mechanism to request an attestation document \textbf{AT} over the booted image \textbf{PCR} values (including both the Enclave Client and the sandbox image), the initial input measurement \textbf{CT}, and the artifact hash \textbf{A}.
The response \textbf{AT} is then shared with the sandbox, so that the build process can include it with the published artifact, published to the transparency log.
Together with proper verification by the client this mitigates \ref{threat:break-verification}.

Importantly, the transparency log ensures that revocation notices (e.g., after discovering hardware vulnerabilities) are visible to all users.
By requiring up-to-date inclusion proofs for artifacts, the end consumer can efficiently verify that they still considered secure.
As such, it lessens the impact of \ref{threat:backdoor-in-repo} and \ref{threat:compromise-hardware}.
Furthermore, transparency logs allow the developer to monitor for leaked signing keys.
They assure users that observed rotations of signing keys are intentional as they know that developers are being notified about them as well.
This mitigates \ref{threat:repository-spoofing}.

After completion, the enclave is destroyed.
This makes the build process stateless which simplifies debugging and reasoning about its life cycle and helps in mitigating \ref{threat:bad-build}, \ref{threat:compromise-hardware}.
Its stateless nature and the clear control of the ingoing code and build instructions ensures that the build is hermetic, i.e. the build cannot accidentally rely on unintended environmental information.
Note that the main build process generally does not require any modifications if it already works with a compatible build runner---it is simply being executed in a sandbox inside an integrity-protected environment.
The developer will only need to add a final step to communicate the artifact path and receive the attestation document \textbf{AD}.

\subsection{Composing A-Bs and R-Bs}
\label{sec:stamp:composing}

We believe that combining our A-Bs and classic R-Bs improves build ergonomics and increases trust.
R-Bs can easily consume A-B artifacts and commit to a hash of the artifact similar to lockfiles that are already used by dependency managers such as Rust's cargo and JavaScript's NPM.
Similarly, A-Bs can consume R-B artifact and even be independent R-B builders themselves.
Due to the attested and controlled environment, existing R-B projects might be able to rely on fewer independent builders when A-Bs are used.

\begin{figure}[b]
    \centering
    \includegraphics[width=0.8\linewidth]{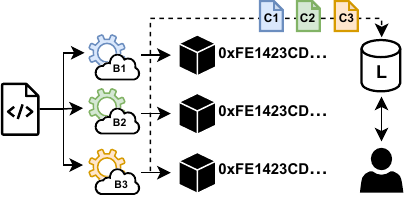}
    \caption{Three attestable builders using different hardware vendors (e.g., Intel, AMD, Arm) perform the same R-B resulting in identical artifacts. The user is then hedged against up to two backdoored TEEs (\S\ref{sec:stamp:composing}).}
    \label{fig:illustration-composition}
\end{figure}

This allows for a setup where the independent builders of an R-B project are distributed across attestable builders running on machines using hardware from different Confidential Computing vendors (see Figure~\ref{fig:illustration-composition}).
In this setting, the guarantees of the R-B imply an anytrust model that is easily verified.
The verifier can use the log to ensure they get a correct build as long as they trust at least one of the Confidential Computing vendors---without having to decide which one.
The reader might find it interesting to compare this with how anonymity networks like mix nets and Tor work where traffic is routed through multiple hops and the unlinkability property holds as long as one of them is trusted.

The trust of A-Bs depends on the trust of their build image.
While the final artifact (or rather its measurement) is attested to and included in the certificate, we rely on the initial image of the machine embedded in the TEE to ensure the correct and secure execution of the build instructions of the source code snapshot.
We believe that R-Bs are important for bootstrapping an A-B system.
Even where the base image can be produced using A-Bs, the very first image should be created using R-Bs and bootstrapped from as little code as possible.
Projects like Bootstrappable Build~\cite{bootstrappable} lay the foundation for this approach.
In the long run, these R-Bs can be executed by attestable builders as described above.

\subsection{Build dependencies}
\label{sec:stamp:architecture:build-dependencies}

The final artifact relies on a number of components that form the Trusted Computing Base (TCB) or are simply direct dependencies specified through the build configuration, e.g., the compilation tool-chain and software libraries.
The TCB also comprises the design of the secure hardware, its firmware, the base image, our implementation of the enclave client, and the sandbox.

Similar to software library dependencies, compiler tool-chains are critical to the trustworthiness of the resulting artifact.
While there is no widely-available support for enforcing particular tool-chains, we side-step this issue by making the TEE image of our prototype implementation a single large build dependency that includes the compilation tool-chains.
Hence the PCR0 measurement covers the compiler tool-chain as well.
For a production system, the large image can be made modular with the compilation tool-chain specified in the build configuration.
While creating a fully verified TEE image is primarily an engineering concern, we demonstrate that A-Bs can build some of its critical components such as the Linux kernel and the Clang compiler (\S\ref{sec:practicalevaluation}).

The attestation document \textbf{AT} can include a reference and cryptographic hash to a full Software Bill of Materials (SBOM).
Including attestation documents for the individual components in such SBOM document, allows consumers to fully capture the impact of known CVEs against the involved components including the chip firmware and the software running in the enclave.
The in-toto standard~\cite{intoto} could be extended for this purpose.

A-Bs do not require the build process to be hermetic, i.e. to be executed without access to the Internet.
As long as dependencies are ``pinned'' using wide-spread support in build tools, e.g. Rust's \texttt{Cargo.lock} file, the primary risks of modification of build assets during transmission (\ref{threat:opa}) are mitigated.
Nevertheless, it is considered best-practice for many large software projects to ``vendor in'' dependencies, i.e. to copy their source code and/or assets into the main repository.
This ensures availability and allows for hermetic builds without access to the Internet during the build step.
If, optionally, confidentiality of the source code and build configuration is required, the developer might prefer hermetic builds to minimize the risk of leakage.
A-Bs are compatible with both pinned dependencies and hermetic builds.

\subsection{TEE attacks and their impact on A-Bs}
\label{sec:stamp:architecture:tee-attacks-and-impact}

No technology provides absolute security, and CC is no different.
There have been recent attacks on popular TEE technologies, including AMD SEV-SNP, that also break integrity properties (\S\ref{sec:relatedwork:attacks}).
However, we find that vulnerabilities affecting TEEs are generally fixed by the chip vendor promptly through firmware updates. Since firmware versions are part of the attestation measurements, the end-user can reject artifacts built in (retrospectively) insecure environments.
To our knowledge, there are no successful attacks against AWS Nitro Enclaves, albeit this might be due to the hardware being less accessible to researchers.

A-Bs only require strong remote attestation and integrity, while confidentiality is optional (\S\ref{sec:stamp:threatmodel:confcomp}).
Importantly, attacks on confidentiality and integrity differ in one important aspect: integrity attacks are inherently active attacks which in turn implies limitations to the window of opportunity.
Therefore, attacks on A-Bs are more difficult for adversaries to achieve since they require persistent access across the fleet and are less opportunistic.

As long as the remote attestation of the most critical measurements, including the CPU firmware, cannot be forged, artifacts that have been built on potentially vulnerable systems can be revoked and rebuilt after the firmware has been patched.
In a larger ecosystem, this might trigger rebuilds of large sub-trees of the dependency graph when far-reaching vulnerabilities in core components are discovered.
If a TEE ever breaks completely, e.g. the internal signing key for a given model can be extracted, all artifacts from such machines can be revoked and rebuilt with a newer generation of TEEs.

\section{Practical evaluation}
\label{sec:practicalevaluation}
We implemented the A-B architecture (\S\ref{sec:stamp:architecture}) to demonstrate its feasibility and to practically evaluate its performance overhead.

\subsection{Implementation}
\label{sec:eval:implementation}

Our prototype uses AWS Nitro Enclaves~\cite{aws_nitro_enclaves} as the underlying Confidential Computing technology due the availability of accessible tooling.
However, it is also possible to achieve similar guarantees with other technologies.
For instance, AMD SEV-SNP might offer security benefits due to a smaller Trusted Computing Base (TCB) and we leave this as an engineering challenge for future work.

AWS Nitro Enclaves are started from EC2 host instances and provide hardware-backed isolation from both the host operation system and the hypervisor through the use of dedicated Nitro Cards.
These cards assign each enclave dedicated resources such as main memory and CPU cores that are then no longer accessible to the rest of the system.
Enclaves boot a \texttt{.eif} image that can be generated from Docker images.
Creation of these images yields PCR0-2\footnote{In the AWS Nitro architecture the values PCR0, PCR1, and PCR2 cover the entire \texttt{.eif} image and can be computed during its build process.} values that can later be attested to.

Since enclaves do not have direct access to other hardware, such as networking devices, all communication has to be done via \textit{vsock} sockets that leverage shared memory.
These provide bi-directional channels that we use to (a) exchange application layer messages between the instance manager and enclave client and (b) tunnel TCP/IP access for the build runner to the code repository.

We implemented two sandbox variants using the lightweight container runtime \textit{containerd} and the hardened \textit{gVisor}~\cite{gvisor2025} runtime which has a compatible API.
Parameters for the sandbox, such as the short-lived authentication tokens for accessing the repository, are passed as environment variables.
Internet access is mediated via Linux network namespaces and results are communicated via a shared log file.
We pass only limited capabilities to the sandbox and the runtime immediately drops the execution context to an unprivileged user.
gVisor provides additional guarantees by intercepting all system calls.
Optionally, this setup can be further hardened using SELinux, seccomp-bpf, and similar.

\begin{figure*}[t]
    \centering
    \includegraphics[width=\textwidth]{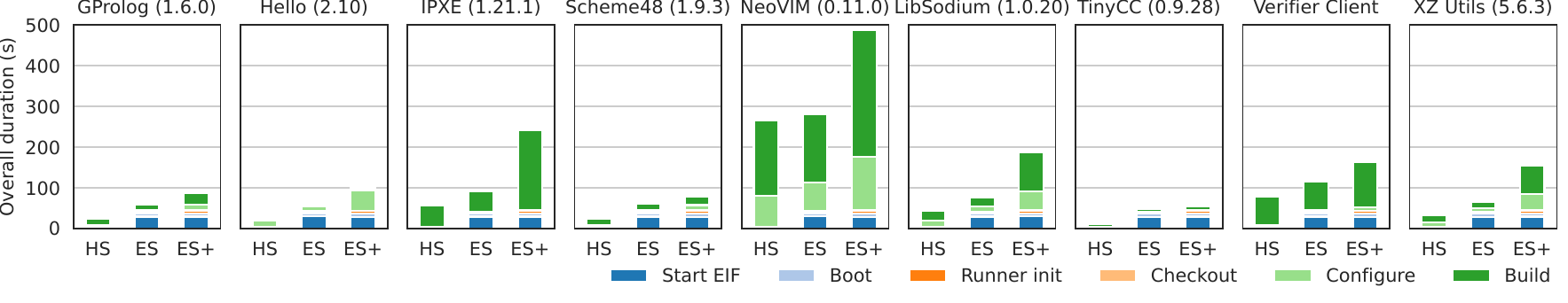}
    \caption{The duration of individual steps for the evaluated projects including the five unreproducible Debian packages and other artifacts. \emph{HS} represents the baseline with a sandbox running directly on the host, \emph{ES} (using containerd) and \emph{ES+} (using gVisor) are variants of our A-B prototype executing a sandboxed runner within an enclave.}
    \label{fig:plot-all-durations}
\end{figure*}

\begin{figure}[t]
    \centering
    \includegraphics[width=\columnwidth]{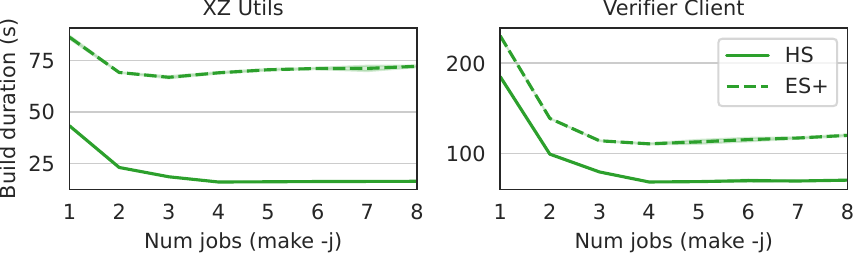}
    \caption{Impact of number of jobs for \texttt{make -j} (left) and \texttt{cargo build -j} (right) with 4 available CPUs.}
    \label{fig:plot-num-jobs}
\end{figure}

\begin{figure}[t]
    \centering
    \includegraphics[width=\columnwidth]{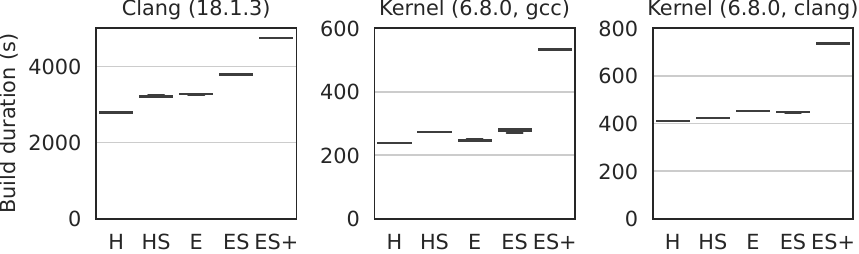}
    \caption{The complex targets \textit{clang} and \textit{kernel} are additionally built without sandboxes on the host \textit{H} and enclave \textit{E}.}
    \label{fig:plot-complex-build-duration}
\end{figure}

As we want to demonstrate ease-of-adoption, we integrated with GitHub Actions.
The Instance Manager exposes a webhook to learn about newly scheduled build workflows and short-lived credentials are acquired using narrowly-scoped personal access tokens (PAT).
Inside the sandbox runs an unmodified GitHub Action Runner (v2.232.0) that is provided by GitHub for self-hosted build platforms.
As such, developers only need to export a PAT, add our webhook, and perform minor edits in their \texttt{.yml} files
\iflongversion (see Appendix~\ref{appendix:example-github-action-file}) \fi
which include updating the runner name and calling the attestation script.

Most components are written in Rust and we leverage its safety features to minimize the overall attack surfaces and avoid logic errors, e.g., through the use of Typestate Patterns~\cite{rust_typestate_pattern}.
Our implementation consists of less than 5\,000 lines of open source code and is available at: \url{https://github.com/lambdapioneer/attestable-builds}.

\subsection{Build targets}
\label{sec:eval:targets}
We demonstrate the feasibility of the A-B approach by building software that appears to be challenging.
First, we build five of the still unreproducible Debian packages.
We start with a list of all unreproducible packages, choose the ones with the fewest but at least two dependencies (to rule out trivial packages), and then use \texttt{apt-rdepends -r} to identify those with the most reverse dependencies, i.e. which likely have a large impact on the build graph.
In addition, we add one with more dependencies.
This results in the following five packages: \emph{ipxe}, \emph{hello}, \emph{gprolog}, \emph{scheme48}, and \emph{neovim}.
Second, we build large software projects including the Linux Kernel (\emph{kernel}, 6.8.0, default config) and the LLVM Clang (\emph{clang}, 18.1.3).
These show that our A-Bs can accommodate complex builds and these two artifacts are also essential for later bootstrapping the base image itself, as these are the versions used in Ubuntu 24.04.
Finally, we augment this set by including \emph{tinyCC} (a bootstrappable C compiler), \emph{libsodium} (a popular cryptographic library), \emph{xz-utils}, and our own verifier client.

For reproducibility, we include copies of the source code and build instructions in a secondary repository with separate branches for each project.
The C-based projects follow a classic configure and make approach and the Rust-based projects download dependencies during the configuration step.

\subsection{Measurements}
\label{sec:eval:measurements}

We build most targets on \emph{m5a.2xlarge} EC2 instances (8\,vCPUs, 32\,GiB).
However, for \emph{kernel} and \emph{clang} we use \emph{m5a.8xlarge} EC2 instances (32\,vCPUs, 128\,GiB).
To allow fair comparison between executions inside and outside the enclave, we assign half the CPUs and memory to the enclave.
At time of writing, the \emph{m5a.2xlarge} instances cost around \$0.34 per hour\footnote{For comparison: at the time of writing, the 4-core Linux runner offered by GitHub costs \$0.016 per minute (\$0.96 per hour).}.
We minimize the impact of I/O bottlenecks by increasing the underlying storage limits to 1000\,MiB/s and 10\,000\,operations/s which incurs extra charges.

In order to better understand how the enclave and the sandbox implementations impact performance, we repeat our experiments across three configurations.
The \textit{host-sandbox (HS)} configuration runs the GitHub Runner using \textit{containerd} on the host and serves as baseline representing a self-hosted build server.
We evaluate two A-B compatible configurations: the \textit{enclave-sandbox (ES)} variant uses the standard \textit{containerd} runtime and the hardened \textit{enclave-sandbox-plus (ES+)} variant uses \textit{gVisor}.
For \emph{kernel} and \emph{clang} we additionally include \textit{H} and \textit{E} configurations without sandboxes.

We are interested in the impact of A-Bs on the duration of typical CI tasks.
For this we have instrumented our components to add timestamps to a log file.
We extract the following steps: 
\textit{Start EIF} allocates the TEE and loads the \texttt{.eif} file into the enclave memory;
then the \textit{Boot} process starts this image inside the TEE;
subsequently the \textit{Runner init} connects to GitHub and performs the source code \textit{Checkout};
finally, the build file performs first a \textit{Configure} step and then executes the \textit{Build}.
We run each combination of build target and configuration three times and report the average.

Figure~\ref{fig:plot-all-durations} plots these durations for the unreproducible Debian packages and the additional targets that we have picked~(\S\ref{sec:eval:targets}).
\iflongversion See Appendix~\ref{appendix:data} for Table~\ref{tab:full-table:main} which contains all measurements (also for other configurations). \fi 
For small builds, the overall duration is dominated by the time required to start and boot the enclave.
Together these two steps typically take around 37.6\,seconds for our 1\,473\,MiB \textit{.eif} file.
These start-up costs can be mitigated by pre-warming enclaves (\S\ref{sec:discussion}).

For small targets we found that the build duration effectively decreases between \textit{HS} and \textit{ES} configurations.
For instance, the NeoVIM build duration (the green bars in Figure~\ref{fig:plot-all-durations}) drop from 184.9\,s (HS) to 167.3\,s (ES, -10\% over HS).
We believe that the enclave is faster because it entirely in memory and therefore mimicks a RAM-disk mounted build with high I/O performance.
Again, \textit{gVisor} (ES+) has a large impact and can increase the build times significantly, e.g., NeoVIM takes 311.7\,s (ES+, +69\% over HS).

The costs for initializing the build runner and checking out the source code are typically less than 9\,seconds overall.
Even though all IP traffic is tunneled via shared memory using \textit{vsock}, the difference between host-based and enclave-based configurations is small.
In fact, for large projects the check-out times sometimes even drops, e.g., \textit{clang} from 148.0\,s (HS) to 117.2\,s (ES).
We believe that the involved Git operations become I/O bound at this size.
However, using \textit{gVisor} (ES+) imposes a overhead for the checkout of up-to 2\,s for small targets and the checkout of the large \textit{clang} target increases from 117.2\,s (ES) to 132.8\,s (ES+).

We found that the impact of \textit{gVisor} (ES+) can be lessened by using parallelized builds, e.g., passing the \texttt{-j} argument to \texttt{make}.
Figure~\ref{fig:plot-num-jobs} shows that ideal number is close to the number of available CPUs.
In our case: 4.
And while increasing numbers past this point is fine for host-based executions, it has negative impact for \textit{ES+}.
\iflongversion See Table~\ref{tab:full-table:scalar-xz}--\ref{tab:full-table:scalar-vc} in Appendix~\ref{appendix:data} for more detailed measurements. \fi

Finally, we build our complex targets \textit{clang} and \textit{kernel} on the larger instance where the TEE is assigned 16\,vCPUs and 64\,GiB.
The larger memory allocation for the TEE increases the \textit{Start EIF} duration from 29.5\,s to 46.4\,s compared to the smaller instance.
Figure~\ref{fig:plot-complex-build-duration} shows that there is also a pronounced impact on the build duration.
For example, \textit{clang}'s build time increased from 54\,minutes (HS) to 63\,minutes (ES, +18\%) or 79\,minutes (ES+, +48\%).

For our overall overhead numbers we build all nine small targets and the two large targets back-back.
With the baseline configuration \textit{HS} this takes 1h22m.
For A-Bs this increases to 1h34m (ES, +14\%) and 2h14m (ES+, +62\%).
These numbers exclude the average start and boot overhead of 42.1s.

\section{Formal verification using \textsc{Tamarin}}
\label{sec:formalverification}

\begin{figure*}[t]
    \centering
    \includegraphics[width=0.97\textwidth]{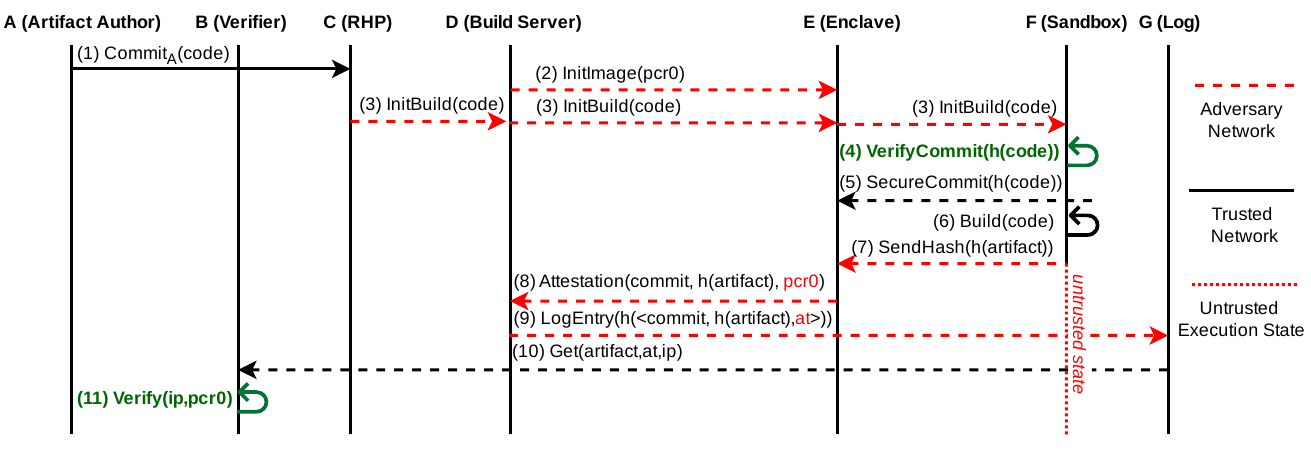}    
    \caption{Protocol flow overview of the formal model, illustrating the interactions and data exchanges between system components and adversary channels. }
    \label{fig:protocol-flow}
\end{figure*}

We use \textsc{Tamarin}~\cite{tamarin_2025}, a security protocol verification tool, to formally model and verify the underlying protocol of A-Bs.
In \textsc{Tamarin}, \emph{facts} represent states of a party involved in a protocol. 
Thus, we can use facts to describe how the components of our system can interact with each other. 
\textsc{Tamarin} allows two types of facts: a linear fact that can be consumed only once as it contributes to the system state, and a persistent fact that can be consumed multiple times. 
A fact in \textsc{Tamarin} is written in the form of \(F(t_1..t_n)\), where \(F\) is the name of the fact and \(t_i\) the value of the current state. 
We use some built-in facts in \textsc{Tamarin}, like \(Fr(x)\), \(In(..)\), and \(Out(..)\).
The \(Fr(x)\) fact generates a fresh random value and the \(In(..)\) and \(Out(..)\) facts are used to receive and send something from and to an adversary-controlled network, respectively.  

\textsc{Tamarin} uses multiset rewriting rules (MSR) to describe state transitions. 
A MSR consists of a name, a left-hand side, an optional middle part, and a right-hand side.
The left-hand side defines the facts that needs to be present in order to initiate the MSR.
The middle part, called \emph{action fact}, is used to label the specific transition and makes it available for the verification step. 
The right-hand side describes the state(s) of the outcome. 

Finally, we define the security properties to be verified. 
\textsc{Tamarin} uses \emph{lemmas} to verify both the expected behavior of the protocol and the results of state transitions based on the given \emph{action facts}.
Considering the \emph{action facts} including an expected time-dependent relation \textsc{Tamarin} derives traces using first-order logic.  
The results allow \textsc{Tamarin} to search a trace that contradicts the lemma and thus the security property.

\subsection{Security properties}
This section outlines various attack categories on security properties used to verify source-to-binary correspondence, including the authenticity of the repository. 
These attack categories are based on our threat model described in Section~\ref{sec:stamp:threatmodel} and we link each category with the respective threat(s) alongside a reference for clarity.
The underlying trust assumptions of our threat model (\S\ref{sec:stamp:threatmodel:assumptions}) also apply for the formal model.
We model the security properties as formulas in a first-order logic using \textsc{Tamarin} \emph{lemmas}.
To verify both protocol behavior and data integrity we utilize \emph{action facts} in the form \(F(x_1..x_n) \#i\) where \(F\) represents the name of the \emph{action fact}, \(x_1..x_n\) the data, and \(\#i\) the time variable for the execution.
Each subsequent paragraph describes the respective attack category and consists of two proofs: one demonstrating that the specific security property can be successfully compromised when not utilizing A-Bs, and another to ensure that there exists no trace where an adversary would be successful when using A-Bs.
We use the function \(h(..)\), which represents a hash function and variables \(c,ct,a,at,ip\) representing the data: \underline{\textbf{c}}ode, \underline{\textbf{c}}ommi\underline{\textbf{t}}hash, \underline{\textbf{a}}rtifact, \underline{\textbf{at}}testation, and \underline{\textbf{i}}nclusion\underline{\textbf{p}}roof. 
\iflongversion The full lemmas of the security properties described below as well as an example illustration of a detected attack by \textsc{Tamarin} are provided in the Appendix~\ref{appendix:security-properties} for reference. \fi

\paragraph{Code manipulation (\ref{threat:compromise-build-process}, \ref{threat:bad-build}, \ref{threat:compromise-hardware})}
This attack category examines whether an adversary can successfully manipulate code during the build process.
Specifically, this includes compromising code on the build server, attacking shared infrastructure, and considering hardware attacks (assuming the TEE to be trustworthy).
Our formal verification begins with proofing that \textsc{Tamarin} can find a trace where an adversary can compromise code \(c\) when specific verification controls, used to verify the commit hash \(ct\), are not incorporated.
Specifically, this lemma proofs that \(\exists\ c, ct : \neg(h(c) = ct)\). 
For the second proof of this attack category, which includes the verification step, \textsc{Tamarin} does not find any trace where an adversary is able to manipulate code without detection. 
This proof verifies that \(\forall\ c,ct : h(c) = ct\). 

\paragraph{Build asset manipulation (\ref{threat:compromise-build-process}, \ref{threat:bad-build}, \ref{threat:backdoor-in-repo}, \ref{threat:compromise-hardware})}
The attack category examines whether an adversary can successfully manipulate a build asset (e.g., the artifact) including potentially malicious libraries side-loaded from external sources.
\textsc{Tamarin} is able to find a trace where an adversary can successfully compromise a build asset \(a\) when specific verification controls, used to verify the inclusion proof \(ip\), are not incorporated.
Specifically, this lemma proves that \(\exists\ c,ct,at,ip : \neg(h(<ct,h(c),h(c)>) = ip)\).
In case of incorporating the verification of the inclusion proof, provided by the transparency log, based on code sent via the adversary network and the attestation \(at\) provide by the TEE, \textsc{Tamarin} does not find a trace where an adversary can manipulate a build asset without detection. 
Specifically, this lemma proves that \(\forall\ c,ct,a,at,ip : h(c) = ct \wedge h(<ct,h(a),at>) = ip\).

\paragraph{Build infrastructure manipulation (\ref{threat:compromise-build-process}, \ref{threat:bad-build}, \ref{threat:compromise-hardware})}
This attack category focuses on successful attacks in which an adversary is able to compromise the infrastructure environment, i.e. the enclave image.
To model this scenario, we transfer the build image through the adversary network so that the adversary can modify it. This analogously covers physical attacks against the machine running the image in an enclave.
Thus, our first lemma in this category verifies whether an adversary can provide an attestation document \(at\) based on a compromised build image without using the trusted PCR value \(p\) to verify the attestation. 
Specifically, it proves that \(\exists\ c,ct,a,p,at : \neg(<c,h(a),p>) = at)\).
However, if we include the proper verification in our model, \textsc{Tamarin} does not find any trace where an adversary can use a manipulated build image without detection. 
The respective proof shows that \(\forall\ c,ct,a,p,at : (<c,h(a),p>) = at \).

\paragraph{Repository Spoofing (\ref{threat:repository-spoofing})}
The last attack category is particularly relevant for spoofing attacks with regards to the repository.  
An adversary might be able to spoof the repository and to create a valid inclusion proof for a particular commit hash of this repository.
In this case, a verifier trying to audit the spoofed repository would get a valid inclusion proof. 
The first lemma, used to verify whether an adversary can successfully spoof the repository when not verifying the inclusion proof shows that \(\exists\ c,ct,a,at,ip : \neg(h(<h(c),h(a),at>) = ip) \).
To prevent such spoofing attacks, the artifact author also needs to verify the corresponding inclusion proof according to the trustworthy reference \(r\). 
Thus, the second lemma proves that \(\forall\ c,ct,a,at,ip,r : h(c) = ct \wedge r = ip \).

\section{Related work}
\label{sec:relatedwork}
The challenge of building software artifacts and distributing them in a trustworthy manner has been known for more than 50~years.
A report on the Multics system by the US Air Force from 1974, was one of the first to present the idea of a compiler trap door~\cite{karger2002thirty}.
Ken Thompson popularized the theme of ``Trusting Trust'' in his Turing Award Lecture in 1984---stating that no amount of source code scrutiny can protect against malicious build processes~\cite{thompson1984trustingtrust}.
In his examples he discusses the implication of a malicious compiler that can introduce a vulnerability in a targeted output binary and preserves this behavior even when it compiles itself from clean source code.
David Wheeler suggests Diverse Double-Compiling (DDC) as a practical solution where one uses a trusted compiler to verify the truthful recompilation of the main compiler~\cite{wheeler2005countering}.
However, this leaves open the question on how to arrive at such a trusted compiler as well as to ensure a trustworthy environment to run the proposed steps in.
Projects like Bootstrappable Builds discuss approaches to build modern systems from scratch using minimal pre-compiled inputs~\cite{bootstrappable}.

The trusted compiler issue can be addressed by having R-Bs and relying either on diverse environments under a at-least-one-trusted assumption or trusting the local setup.
The inherent challenges are discussed in academic literature for both individual tools and the overall environment~\cite{de2014challenges,lamb2021reproduciblefoss}.
More papers include industry perspectives on business adoption~\cite{butler2023business}, 
experience reports for large commercial systems~\cite{shi2021experience},
and importance and challenges as perceived by developers~\cite{fourne2023cispa}.
In addition, there has been work aiming at making build environments and tools more deterministic~\cite{navarro2020reproducible,xiong2022towardsjava,glukhova2017tools}.

While deterministic builds aid verification, it also means that the exact same code will be deployed to each target system.
This can help attackers since the context of vulnerable code, e.g. register assignments and code pointers, will be exactly the same for each target---potentially also allowing extensive local experiments in the case of generally available software to fine-tune attacks.
``Software Diversity'' aims at removing this predictability from the generated artifacts by including randomized variation during compilation, linking, and execution stages~\cite{larsen2014softwarediversity}.
A-Bs can support software diversification during the compilation and linking phase since it allows for non-determinism, while R-Bs cannot.
However, all approaches are compatible with run-time diversification techniques such as Address Space Layout Randomization (ASLR).

Similar to our approach of using \textit{Confidential Computing} (CC) for providing integrity, Russinovich et al. introduce the idea of Confidential Computing Proofs (CCP) as a more scalable alternative to Zero Knowledge Proofs which rely on heavy and slow cryptography~\cite{russinovich2024confidential}.
A-Bs can be seen as a form of CCP that is persisted using a transparency log.
Meng et al. propose the use of TPMs in software aggregation to reduce the size of hard-coded lists of trusted binary artifacts~\cite{meng2009remote}, but their work lacks a security model and does not generalize to cloud-based CI/CD with untrusted build processes.
Others also identified the challenges and opportunities of Confidential Computing as a Service (CCaaS) and our deployment model is inspired by the work by Chen et al.~\cite{chen2023verified}.

Trust of pre-built dependencies is key for \textit{supply chain security} and software updates.
The framework Supply-chain Levels for Software Artifacts (SLSA) provides helpful threat-modeling and taxonomy to discuss guarantees provided by different systems~\cite{slsa}.
Both R-Bs and A-Bs could be adopted as a new level L4 (see Table~\ref{tab:slsa-levels}).
Frameworks like SLSA become particularly valuable when integrated with codified descriptions such as the in-toto standard~\cite{intoto}
CHAINIAC demonstrates how to transparently ship updates using skipchains and verified builds~\cite{nikitin2017chainiac}.

\begin{table}[t]
    \centering
    \caption{The existing SLSA levels L0--L3 adapted from~\cite{slsa} and possible new L4 levels for A-Bs and R-Bs.}
    \renewcommand*{\arraystretch}{1.05}
    \begin{tabular}{c p{7.25cm}}
        \toprule
        ~ & Requirements \& focus \\
        \midrule
        \textbf{L4} & Attestable build \\ ~ & $\rightharpoonup$ Attested trust in builder \\
        \textbf{L4} & Reproducible build \\ ~ & $\rightharpoonup$ Verifiable trust in builder \\
        \midrule
        \textbf{L3} & Hardened build platform \\ ~ & $\rightharpoonup$ Tampering during the build \\
        \textbf{L2} & Signed provenance from a hosted build platform \\ ~ & $\rightharpoonup$ Tampering after the build \\
        \textbf{L1} & Provenance showing how the package was built\\ ~  & $\rightharpoonup$ Mistakes, documentation \\
        \textbf{L0} & n/a \\
        \bottomrule
    \end{tabular}
    \label{tab:slsa-levels}
\end{table}

\textit{Sigstore} provides an ecosystem~\cite{newman_2022} to sign and verify artifacts. 
The authentication certificate together with the artifact hash and the signature is then logged in a transparency log, allowing later verification of a downloaded artifact. 
Both A-B and the Sigstore project incorporate a transparency log for end-to-end verification.
Sigstore makes the signature process verifiable, while we use a transparency log to store metadata about the attested build. 

\label{sec:relatedwork:attacks}
No technology provides absolute security and, in recent years, various researchers have been able to break the security guarantees of TEEs. 
 Since A-Bs do not require confidentiality, many attacks~\cite{ciphertextsidechannel22,morbitzer2018severed,cachewarp24,li2021cipherleaks,gotzfried2017cache,lee2017inferring,chen2019sgxpectre,wilke24,counterseveillance25}, including classical side-channel attacks, do not affect its security properties. 
However, A-Bs rely on strong integrity protection provided by the underlying TEE. 
To the best of our knowledge, there are no successful attacks that compromise the integrity guarantees of AWS Nitro Enclaves. 
\iflongversion We discuss three recent attacks against comparable TEEs in Appendix~\ref{appendix:attacks}, such as AMD SEV-SNP and Intel TDX, that stand in as representative attacks against typical CC technology. \fi

\section{Deployment consideration}
\label{sec:discussion}
\paragraph{Going beyond executable binaries}
In this paper we focus on executable binary artifacts that are given to verifiers, e.g., a user downloading new software from the Internet.
However, we can also attest other build process outputs.
One natural area are supply-chains of software libraries.
In such a system, each dependency is built in an attestable manner and the downstream builders verify each included dependency.
Since this verification step is part of the attested build process, trust spreads transitively.
A-Bs can also attest non-binary artifacts.
Examples include results of a vulnerability scanning program (an artifact is secure) or accuracy scores of a benchmark that is run in CI against the built artifact (an artifact meets a certain standard).
Another compelling application of the A-B paradigm is its use as part of an issuing authority, e.g., an SSL provider who needs to perform certain checks while creating a new certificate, where trust is an essential aspect.

\paragraph{Integrating with existing CI/CD systems.}
Our prototype already integrates with the GitHub Actions CI/CD product using \textit{workflow files} (\texttt{.yml}).
We found that the required changes are typically less than 10 lines and \iflongversion Appendix~\ref{appendix:example-github-action-file} \fi shows a side-by-side comparison of the changes to a typical workflow file.
Overall, the developer experience remains the same.
\iflongversion Figure~\ref{fig:screenshot} in Appendix~\ref{appendix:figures} shows a web screenshot during our evaluation. \fi
A-Bs can be provided by a third-party providing audited base images and run on untrusted CSPs.

\paragraph{Mitigating performance impact}
In our evaluation, A-Bs incur a large start-up overhead.
However, in practice this can be mitigated by maintaining a number of ``pre-warmed'' enclaves that are booted, but have not yet fetched any source code.
Additionally, as EC2 instances can host a mix of multiple enclaves of various size---given sufficient vCPU and RAM resources---the overall costs can remain low.
A load balancer can then redirect build requests to the most suitable ready instance.

\paragraph{Extending the log} In this paper, our transparency log contains entries that link source code snapshots and binary artifacts.
However, in a production system these logs can be extended with various types of entries that more holistically capture the security of a given artifact.
For example, auditors might provide \textit{SourceAudit} entries signed by their private key to vouch for a given code snapshot and maybe even link them to a set of audit standards published by regulators.
Software and hardware vendors might publish \textit{RevocationNotices} when new vulnerabilities are discovered.
Based on these, the artifact authors can then ask the independent log monitors to regularly provide compact proofs that attest to the fact that (a) an artifact was built from a given code snapshot, (b) that code snapshot was audited to an accepted standard, and (c) there are no revocation notices affecting this version.
The verifier then only needs to check threshold many such up-to-date proofs instead of having to inspect the entire log themselves.

\subsection{Case studies}
\label{sec:discussion:case-studies}

\begin{figure}
    \centering
    \includegraphics[width=0.7\linewidth]{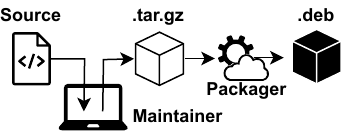}
    \caption{Illustration of the XZ build chain.}
    \label{fig:illustration-xz}
\end{figure}

A key aspect of the XZ incident (CVE-2024-3094)~\cite{lins2024libxz} was that the adversary added an additional build asset \texttt{build-to-host.m4} to the tarball used by the packager to build the final artifact (see Figure~\ref{fig:illustration-xz}).
Having some pre-generated files (e.g., configure script) is common for projects utilizing \textit{Autoconf} to make the build process easier for others. 
However, as these build assets are not included in the repository, it is difficult to verify whether they have been generated in a trustworthy manner. 
Additionally, the concept of R-Bs may not apply as the resulting artifact likely differs when built on another build host. 
We believe that A-Bs can add an extra layer of transparency by allowing to verify that the build assets were created in a trustworthy environment based on a specific source code snapshot.
Thus, using A-Bs with XZ compels the adversary to use a repository containing all required source code, including the covert \texttt{build-to-host.m4} file, to create the tarball that is finally used by the packager. 

The SolarWinds hack~\cite{wired2023solarwind} had a large impact after adversaries successfully compromised a critical supply-chain by implanting a backdoor in a critical software package.
The defining aspect of this episode was that the adversaries did not modify the source code in the repository, but were able to compromise the build infrastructure (\ref{threat:compromise-build-process}) in a covert manner. 
Specifically, SUNSPOT was used to inject a SUNBURST backdoor into the final artifact by replacing the corresponding source file during the build process~\cite{sunspot2021solarwind}.
If A-Bs were used, the change in the PCR values or a failing attestation would have indicated that the build image was modified.

These deployment considerations and potential mitigation for such supply-chain attacks are particularly important for audited, but closed-source firmware.
A practical attack demonstration where the authors explain how to engineer a backdoored bitcoin wallet highlights this issue for high-assurance use-cases~\cite{scott_2024}.
We believe that A-Bs can help mitigate such attacks, as the build step itself runs within a trusted and verifiable environment, thus preventing persistent and covert compromise.

\section{Conclusion}
\label{sec:conclusion}
We presented Attestable Builds (A-Bs) as a new paradigm to provide strong source-to-binary correspondence in software artifacts.
Our approach ensures that a third-party can verify that a specific source-code snapshot was used to build a given artifact.
A-Bs take into account the modern reality of software development which often relies on a large set of third-parties and cloud-hosted services.
We demonstrated this by integrating our prototype with a popular CI/CD framework as part of our evaluation.

Our prototype builds existing projects with no source code changes, and only minimal changes to existing build configurations.
We show that it has acceptable overhead for small projects and can also build notoriously complex projects such as LLVM clang.
More interesting use-cases are possible, such as attesting non-binary artifacts and building composite systems which also support reproducible builds.
Importantly, A-Bs can be pragmatically adopted for difficult gaps in R-B projects as well as an off-the-shelf solution for migrating entire projects.

\section*{Acknowledgments}
We thank the reviewers and shepherd for their feedback.
We also thank Jenny Blessing, Adrien Ghosn, Alberto Sonnino, and Tom Sutcliffe for the valuable discussions and feedback on earlier versions of this paper.
All errors remain our own.
Daniel is supported by Nokia Bell Labs.
This work has been carried out within the scope of Digidow, the Christian Doppler Laboratory for Private Digital Authentication in the Physical World and has partially been supported by the LIT Secure and Correct Systems Lab.
We gratefully acknowledge financial support by the Austrian Federal Ministry of Labour and Economy, the National Foundation for Research, Technology and Development, the Christian Doppler Research Association, 3 Banken IT GmbH, ekey biometric systems GmbH, Kepler Universitätsklinikum GmbH, NXP Semiconductors Austria GmbH \& Co KG, Österreichische Staatsdruckerei GmbH, and the State of Upper Austria.

\bibliographystyle{ACM-Reference-Format}
\bibliography{references}

\iflongversion 

\newpage
\appendix
\label{appendix}
\section{Additional figures}
\label{appendix:figures}

\begin{figure}[h!]
    \centering
    \includegraphics[width=0.99\columnwidth]{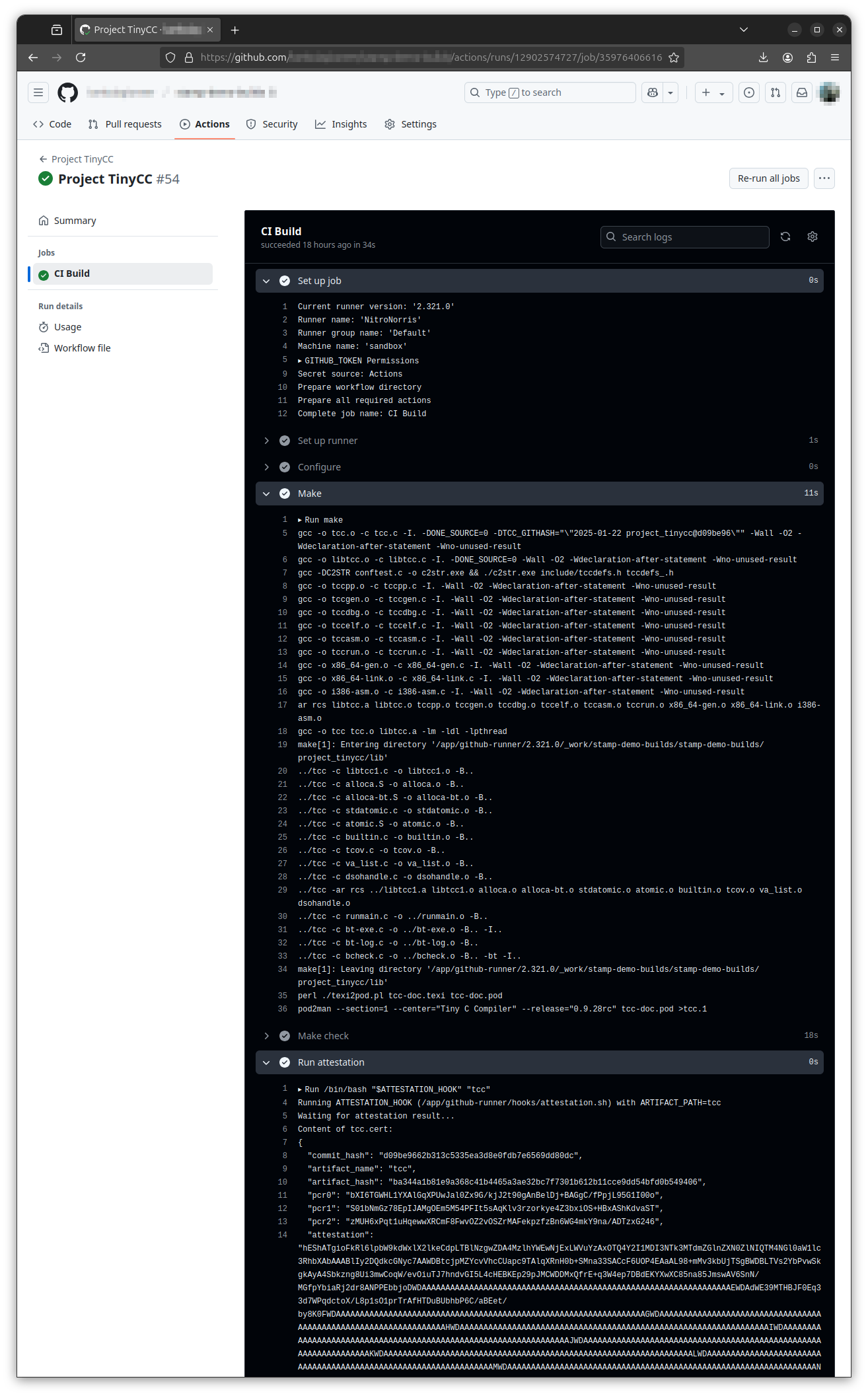}
    \caption{Screenshot of the GitHub Action CI running our prototype. The first section shows our runner configuration. The middle section the output from the main build step. The bottom section shows a report from the generated certification file including the PCR0--2 values, the source code commit hash, the artifact hash, and the signed attestation document.}
    \label{fig:screenshot}
\end{figure}

\newpage
\section{Attacks against Confidential Computing}
\label{appendix:attacks}

In 2025, De Meulemeester, Wilke, et al. published the BadRAM attack which undermines integrity guarantees of AMD SEV-SNP \cite{badramsp25}.
Using the BadRAM attack, an adversary manipulates the Serial Presence Detect (SPD) chip which is used to communicate memory properties to the BIOS.
In particular, an adversary can manipulate the ciphertext and the reverse map table data structure of AMD SEV-SNP hardware.
Normally, this requires physical access.\footnote{Such vulnerabilities relying on on-line DRAM integrity would have been outside AMD's scope: ``These attacks are very complex and require a significant level of local access and resources to perform''~\cite{amdsevsnpwhitepaper20}}
However, the researchers identified certain DRAM vendors that did not lock the SPD chip and thus allowed adversaries to perform the attack remotely through the host system. 
In addition, BadRAM enables an adversary to present the attestation digest of one enclave image while executing another.
However, it does not break the most important part of the attestation: the firmware measurements. 
As such, recipients of the attestation document can distrust any builds produced by systems running firmware versions with known vulnerabilities, thus mitigating the impact of this attack on A-Bs.
The BadRAM vulnerability was disclosed to AMD (CVE-2024-21944) and was mitigated through a firmware update~\cite{amdbadram24}.

Two additional attacks compromising integrity are \textsc{wesee}~\cite{wesee24} and \textsc{heckler}~\cite{schluter2024heckler}. 
In \textsc{wesee} (CVE-2024-25742), Schlüter et al. break AMD SEV-SNP by interrupt injections allowing an adversary to compromise the confidentiality and integrity of a VM.
The \textsc{heckler} (CVE-2024-25744, CVE-2024-25743) attack, similarly, uses adversary-controlled non-timer interrupts to break both the confidentiality and integrity guarantees of both AMD SEV-SNP and Intel TDX.
Both attacks have been mitigated by kernel security patches~\cite{mitigationinterrupt23,kernelpatchsevsnp24,debianpatchsevsnp24}.

Overall, the number of CVEs related to AMD SEV(-ES/SNP) and Intel TDX are low despite active research in this area. 
Misono et al. found that there are 17 known firmware bugs, 3 hardware vulnerabilities, and 4 design issues from 2019 to June 2024 for the Host-to-Guest attack type on AMD SEV-SNP~\cite{misono24}.
All of them have been mitigated, e.g., through firmware updates.

\clearpage
\section{GitHub Action integration}
\label{appendix:example-github-action-file}

We show the required modification of an exemplary GitHub Action workflow file (Listing~\ref{listing:before}) into one that uses our A-B prototype (Listing~\ref{listing:after}).
In addition, the repository owner needs to provide a Personal Access Token (PAT), so that we can register a new runner.

In the new version, the \texttt{runs-on} field now uses the name of the self-hosted A-B runner.
Also, the repository no longer checkout its source manually, but this is done by the A-B runner before any other build code is being executed to avoid interference with the calculation of the repository commit hash.
Next, we add a call to the attestation service by calling the provided \texttt{ATTESTATION\_HOOK} environment variable which contains the path to the attestation executable provided by the runner.
Optionally, we upload the attestation certificate together with the artifact.
\vfill
\lstset{style=actionstyle}
\begin{lstlisting}[label=listing:before,caption=A typical GitHub Action workflow file for a small Rust project.]
name: CI for a Rust project

on:
  push:
    branches: [ "main" ]
  pull_request:
    branches: [ "main" ]

jobs:
  lint:
    runs-on: ubuntu-24.04
    name: Lint (clippy & fmt)
    steps:
    - name: Check out code
      uses: actions/checkout@v4.2.0
    - name: Lint
      run: cargo clippy --verbose
    - name: Format
      run: cargo fmt -- --check

  build-and-test:
    runs-on: ubuntu-24.04
    name: Build and Test
    steps:
    - name: Check out code
      uses: actions/checkout@v4.2.0
    - name: Build
      run: cargo build --verbose
    - name: Test
      run: cargo test --verbose

      
    - name: Upload artifacts
      uses: actions/upload-artifact@v4.6.0
      with:
        name: artifacts
        path: |
          target/debug/executable
          _
\end{lstlisting}

\newpage
~
\vfill

\lstset{style=actionstyle}
\begin{lstlisting}[label=listing:after,caption=The modified version of the GitHub Action workflow file using our prototype and providing attestation.]
name: CI for a Rust project

on:
  push:
    branches: [ "main" ]
  pull_request:
    branches: [ "main" ]

jobs:
  lint:
    runs-on: attested-build-runner
    name: Lint (clippy & fmt)
    steps:

    
    - name: Lint
      run: cargo clippy --verbose
    - name: Format
      run: cargo fmt -- --check

  build-and-test:
    runs-on: attested-build-runner
    name: Build and Test
    steps:

    
    - name: Build
      run: cargo build --verbose
    - name: Test
      run: cargo test --verbose
    - name: Attestation
      run: $ATTESTATION_HOOK target/debug/executable
    - name: Upload artifacts
      uses: actions/upload-artifact@v4.6.0
      with:
        name: artifacts and certificate
        path: |
          target/debug/executable
          target/debug/executable.cert
\end{lstlisting}

\newpage
\section{Security properties}
\label{appendix:security-properties}
This section outlines the lemmas used for our formal verification. 
We use single letter variable names for better readability and to keep lemmas short.
See Table~\ref{table:notation} for the used notation.

\begin{table}
	\caption{Notations used in \textsc{Tamarin} lemmas}
	\begin{tabular*}{\columnwidth}{ >{\raggedright\arraybackslash}p{20mm} >{\raggedright\arraybackslash}p{58mm} }
		\toprule
		Notation & Description\\
		\midrule		
        \( c,ct,a,at,p,ip \) & c(code), ct(commit), a(artifact), at(attestation), p(pcr0), ip(incl. proof)\\
        \(Commit(c)\) & Artifact author commit code to RHP. \\
        \(Publish(c)\) & Publish code to adversary network. \\
        \(InitImage(c)\) & Initialize image with PCR and publish it via the adversary network. \\
        \(InitBuild(c)\) & Fetch Code from adversary network. \\
        \(Secure\)-\(Commit(ct)\) & Store commit hash before entering untrusted execution state. \\
        \(Commit-\) \(Verify(c,ct)\) & Verify the commit hash (h(c) = ct). \\
        \(Artifact(a)\) & Provide build artifact. \\
        \(Attestation(at)\) & Provide attestation document based on PCR from adversary network. \\
        \(LogEntry(ip)\) & Provide incl. proof of given log entry. \\
        \(LogVerify(f,ip)\) & Verify inclusion proof for given f where f = <ct,h(a),at>\\
        \(ATVerify(f,at)\) & Verify attestation for given f where f = <c,h(a),p>.\\      
		\bottomrule
	\end{tabular*}
	\label{table:notation}
\end{table}

\paragraph{Code manipulation}
The following lemma proofs that the adversary is able to successfully compromise the code without detection when the verification controls are not used. 
\begin{equation*}
\begin{split}
\exists\,c,\,ct,\,\#i,\,\#j.\ 
\left(
\left(\textrm{\(Publish\)}\left(c\right)\ @\,\#i\right) \wedge \right.\\
\left.\left(\textrm{\(SecureCommit\)}\left(ct\right)\ @\,\#j\right) \right) 
\wedge \left(\neg\left(\textrm{h}\left(c\right)=ct\right)\right)
\end{split}
\end{equation*}

This lemma proofs that the adversary is not able to compromise the code without detection when using the specific verification controls. 
\begin{equation*}
\begin{split}
\forall\,c,\,ct,\,\#i,\,\#j,\,\#k,\,\#l.\
\left(\left(\left(
\left(\textrm{\(Commit\)}\left(c\right)\ @\,\#i\right) \wedge \right. \right. \right. \\
\left.\left.\left.\left(\textrm{\(Publish\)}\left(c\right)\ @\,\#j\right)\right) \wedge \left(\textrm{\(SecureCommit\)}\left(ct\right)\ @\,\#k\right) \right) \wedge \right.\\ \left. \left(\textrm{\(CommitVerify\)}\left(c,\,ct\right)\ @\,\#l\right) \Rightarrow \left( h\left(c\right) = ct\right)
\right) 
\end{split}
\end{equation*}

\paragraph{Build asset manipulation}

\begin{equation*}
\begin{split}
\exists\,c,\,ct,\,at,\,ip,\,\#i,\,\#j,\,\#k,\,\#l.\ \left(\left(\left(\left(\left(\textrm{\(Publish\)}\left(c\right)\ @\,\#i\right) \wedge \right.\right.\right.\right. \\
\left.\left.\left(\textrm{\(SecureCommit\)}\left(ct\right)\ @\,\#j\right)\right) \wedge \left(\textrm{Attestation}\left(at\right)\ @\,\#k\right)\right) \wedge \\
\left. \left.
\left(\textrm{\(LogEntry\)}\left(ip\right)\ @\,\#l\right)\right) \wedge \left(\neg \left( h\left(c\right) = ct\right)\right)\right) \wedge \\
\left(\neg\left(h\left(<ct,h\left(c\right),h\left(c\right)>\right) = ip \right)\right)
\end{split}
\end{equation*}

The lemma below verifies the inclusion proof using the action fact \(LogVerify(<ct,h(a),at>,ip)\), and \textsc{Tamarin} does not detect any trace, where such an attack is possible.

\begin{equation*}
\begin{split}
\forall\,c,\,ct,\,a,\,at,\,ip,\,\#i,\,\#j,\,\#k.\
 \\
\left(\left(
\left(\textrm{\(Publish\)}\left(c\right)\ @\,\#i\right) 
\left(\textrm{\(CommitVerify\)}\left(c,\,ct\right)\ @\,\#j\right) \right) \wedge \right.\\
\left. 
\left(\textrm{\(LogVerify\)}\left(<ct,\,h\left(a\right),at>,\,ip\right)\ @\,\#k\right)\right) \\
\Rightarrow \left(\left(h\left(c\right) = ct\right) \wedge \left(h\left(<ct,\,h\left(a\right),\,at>\right) = ip \right)\right)
\end{split}
\end{equation*}

\paragraph{Build infrastructure manipulation}

\begin{equation*}
\begin{split}
\exists\,c,\,ct,\,a,\,p,\,at,\,\#i,\,\#j,\,\#k,\,\#l,\,\#m.\ \\
\left(\left(\left(
\left(
\left(\textrm{\(Publish\)}\left(c\right)\ @\,\#i\right)
\left(\textrm{\(SecureCommit\)}\left(ct\right)\ @\,\#j\right)
\right) \wedge \right.\right.\right.\\
\left.
\left.
\left(\textrm{Artifact}\left(a\right)\ @\,\#k\right)
\right) \wedge 
\left(\textrm{InitImage}\left(p\right)\ @\,\#l\right)
\right) \wedge \\
\left.
\left(\textrm{Attestation}\left(at\right)\ @\,\#m\right)
\right) \wedge
\left(\neg\left(<c,\,h\left(a\right),\,p>=at\right)\right)
\end{split}
\end{equation*}

The following lemma utilizes the verification control \(ATVerify(..)\) provided by Attestable Builds. 
\textsc{Tamarin} cannot find any trace where an adversary is able to successfully compromise the build image without detection.

\begin{equation*}
\begin{split}
\forall\,c,\,ct,\,a,\,p,\,at,\,\#i,\,\#j,\,\#k,\,\#l,\,\#m,\,\#n.\ \\
\left(\left(\left(\left(
\left(
\left(\textrm{\(Publish\)}\left(c\right)\ @\,\#i\right)
\left(\textrm{\(SecureCommit\)}\left(ct\right)\ @\,\#j\right)
\right) \wedge \right.\right.\right.\right.\\
\left.
\left.
\left(\textrm{Artifact}\left(a\right)\ @\,\#k\right)
\right) \wedge 
\left(\textrm{InitImage}\left(p\right)\ @\,\#l\right)
\right) \wedge \\
\left.
\left(\textrm{Attestation}\left(at\right)\ @\,\#m\right)
\right) \wedge \\
\left.
\left(\textrm{\(ATVerify\)}\left(<c,\,h\left(a\right),p>,\,at\right)\ @\,\#n\right)
\right) \\
\Rightarrow
\left(<c,\,h\left(a\right),\,p>=at\right)
\end{split}
\end{equation*}

\paragraph{Repository spoofing}

The following lemma proofs that an adversary would be able to successfully spoof a repository when the corresponding verification control is not involved. 
\begin{equation*}
\begin{split}
\exists\,c,\,ct,\,a,\,at,\,ip,\,\#i,\,\#j,\,\#k,\,\#l,\,\#m,\,\#o.\ \\
\left(\left(\left(\left(\left(
\left(
\left(\textrm{\(Commit\)}\left(c\right)\ @\,\#i\right)
\wedge
\left(\textrm{\(Publish\)}\left(c\right)\ @\,\#j\right)
\right) \wedge \right.\right.\right.\right.\right.\\
\left.
\left(\textrm{\(SecureCommit\)}\left(ct\right)\ @\,\#k\right)
\right) \wedge \\
\left.
\left.
\left(\textrm{\(CommitVerify\)}\left(c,\,ct\right)\ @\,\#l\right)
\right) \wedge 
\left(\textrm{\(Artifact\)}\left(a\right)\ @\,\#m\right)
\right) \wedge \\
\left.
\left.
\left(\textrm{\(Attestation\)}\left(at\right)\ @\,\#n\right)
\right) \wedge 
\left(\textrm{\(LogEntry\)}\left(ip\right)\ @\,\#o\right)
\right) \\
\left(\neg\left(h\left(<h\left(c\right),\,h\left(a\right),\,at>\right)=ip\right)\right)
\end{split}
\end{equation*}

Our model involves a verification step \(RepositoryVerify(..)\) performed by the artifact author to mitigate repository spoofing.
The following lemma proofs that an adversary cannot successfully spoof a repository when using Attestable Builds.

\begin{equation*}
\begin{split}
\forall\,c,\,ct,\,a,\,at,\,ip,\,r,\,\#i,\,\#j,\,\#k,\,\#l,\,\#m,\,\#o,\,\#p.\ \\
\left(\left(\left(\left(\left(\left(
\left(
\left(\textrm{\(Commit\)}\left(c\right)\ @\,\#i\right)
\wedge
\left(\textrm{\(Publish\)}\left(c\right)\ @\,\#j\right)
\right) \wedge \right.\right.\right.\right.\right.\right.\\
\left.
\left(\textrm{\(SecureCommit\)}\left(ct\right)\ @\,\#k\right)
\right) \wedge \\
\left.
\left.
\left(\textrm{\(CommitVerify\)}\left(c,\,ct\right)\ @\,\#l\right)
\right) \wedge 
\left(\textrm{\(Artifact\)}\left(a\right)\ @\,\#m\right)
\right) \wedge \\
\left.
\left.
\left(\textrm{\(Attestation\)}\left(at\right)\ @\,\#n\right)
\right) \wedge 
\left(\textrm{\(LogEntry\)}\left(ip\right)\ @\,\#o\right)
\right) \\
\left.
\left(\textrm{\(RepositoryVerify\)}\left(r,\,ip\right)\ @\,\#p\right)
\right) \Rightarrow
\left(\left(h\left(c\right)=ct\right) \wedge
\left(r=ip\right)\right)
\end{split}
\end{equation*}

\clearpage
\begin{figure*}
  \centering
  \includegraphics[height=0.60\textheight,angle=90]{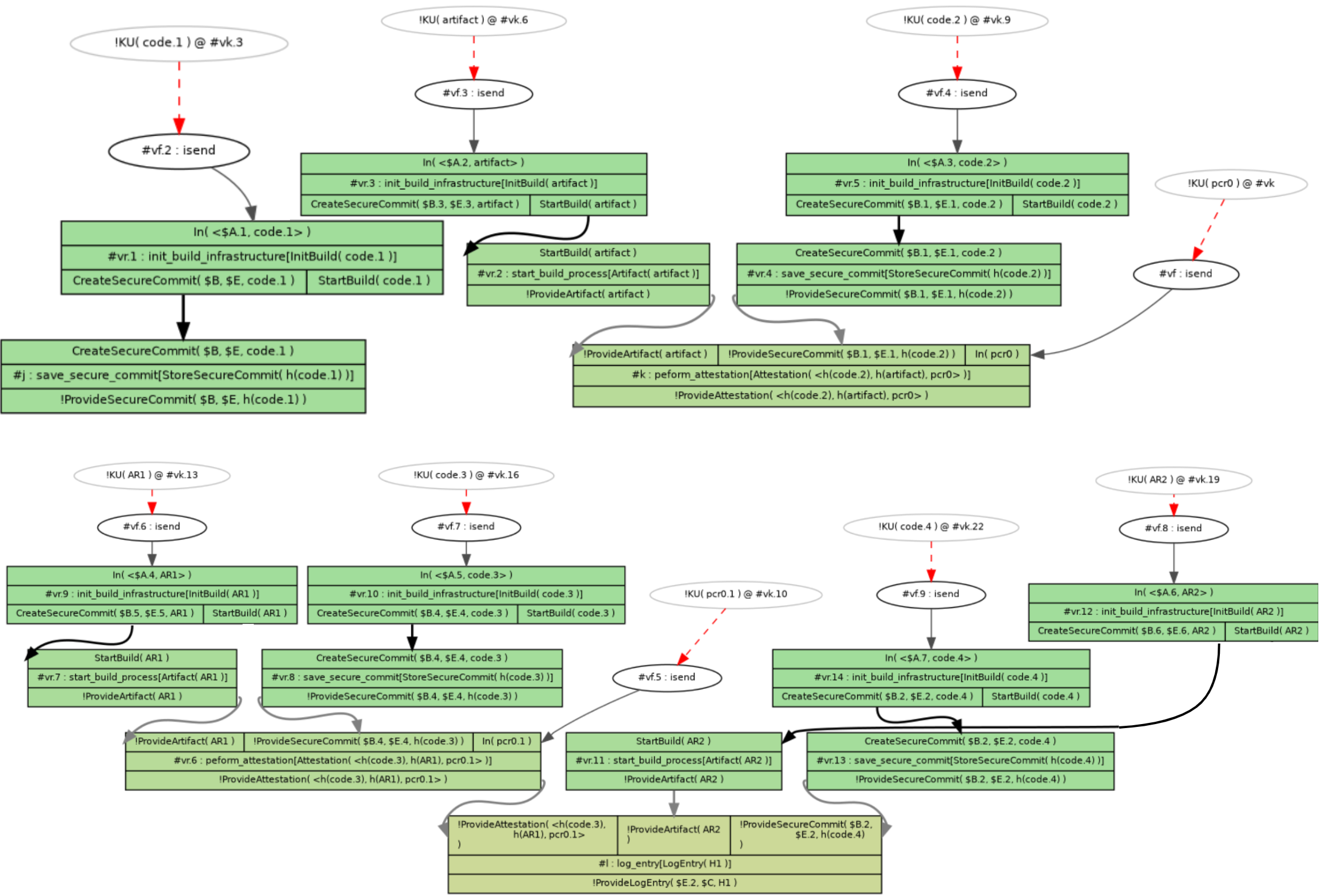}
    \caption{This plot shows initial attack traces that \textsc{Tamarin} found. Traces like these show that our initial adversary assumptions would affect regular build systems. When enabling the A-B constraints, \textsc{Tamarin} fails to find any.}
    \label{fig:tamarin-sanitycheck}
\end{figure*}

\clearpage

\section{Additional evaluation material}
\label{appendix:data}

This appendix includes additional data and plots from our evaluation (\S\ref{sec:practicalevaluation}).
Table~\ref{table:debian-packages} lists the dependencies and reverse dependencies for the unreproducible Debian packages that we have selected for our evaluation.
Table~\ref{tab:full-table:main} shows all individual durations from our main evaluation.
Tables~\ref{tab:full-table:scalar-xz}--\ref{tab:full-table:scalar-vc} show all individual durations from the job number experiments for XZ Utils and Verifier Client respectively.
Figure~\ref{fig:plot-all-durations-tall} is a larger version of Figure~\ref{fig:plot-all-durations}. Figure~\ref{fig:plot-complex-durations-tall} shows the stacked bar chart for the complex targets.

\begin{figure}[h]
    \centering
    \includegraphics[width=\columnwidth]{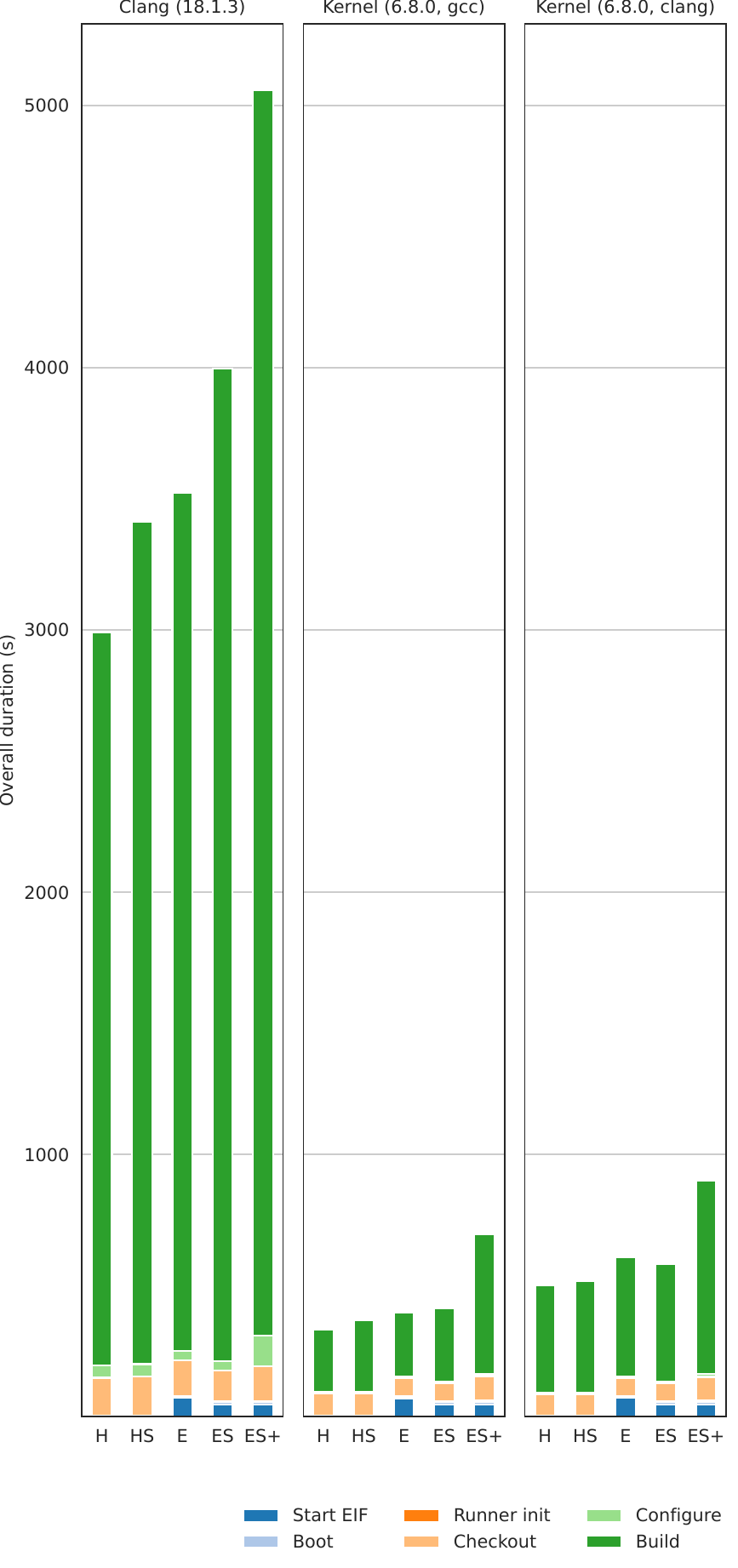}
    \caption{A variant of Figure~\ref{fig:plot-all-durations-tall}. The complex targets \textit{clang} and \textit{kernel} are additionally built without sandboxes on the host \textit{H} and enclave \textit{E}.}
    \label{fig:plot-complex-durations-tall}
\end{figure}

\begin{figure*}[h]
    \centering
    \includegraphics[width=\textwidth]{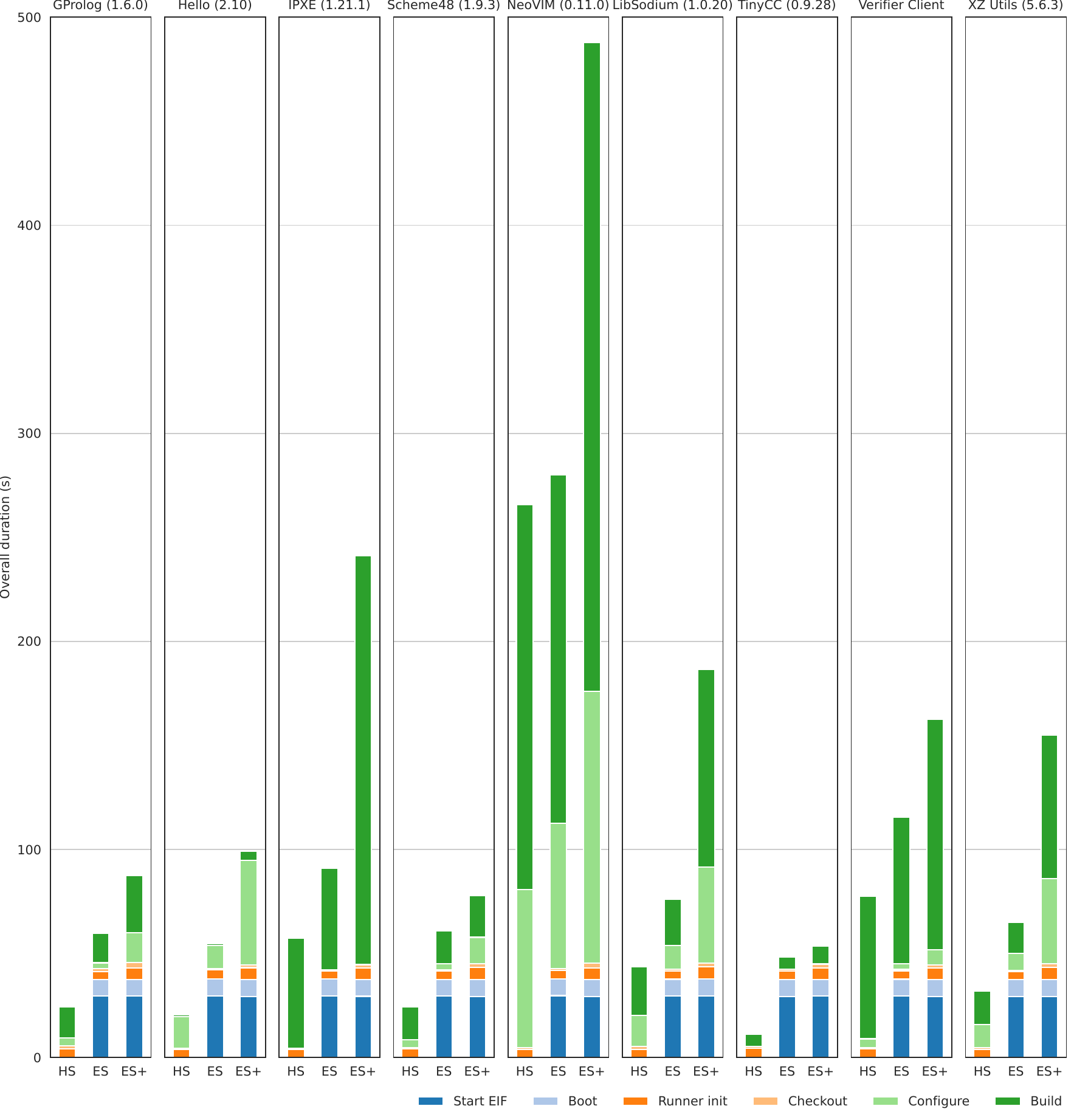}
    \caption{A taller version of Figure~\ref{fig:plot-all-durations}. The duration of individual steps for the evaluated projects including the five unreproducible Debian packages and other artifacts. \emph{HS} represents the baseline with a sandbox running directly on the host, \emph{ES} (using containerd) and \emph{ES+} (using gVisor) are variants of our A-B prototype executing a sandboxed runner within an enclave.}
    \label{fig:plot-all-durations-tall}
\end{figure*}

\begin{table}[h!]
	\caption{Selected unreproducible Debian packages based.}
	\begin{tabular*}{\columnwidth}{ >{\raggedright\arraybackslash}p{22mm} >{\raggedright\arraybackslash}p{26mm}
    >{\raggedright\arraybackslash}p{32mm}}
		\toprule
		Package & Number of Dependencies & Number of Reverse Dependencies \\
		\midrule		
        ipxe & 3 & 2 \\
        hello & 4 & 3 \\
        gprolog & 4 & 2 \\
        scheme48 & 4 & 2 \\
        neovim & 16 & 39  \\        
		\bottomrule
	\end{tabular*}
	\label{table:debian-packages}
\end{table}

\begin{table*}[h]
\caption{Build times for all targets and modes. Missing items indicate that they did not work out-of-the-box, e.g., because Amazon Linux 2023 is missing some dependencies in H mode. All values in seconds and averaged over three iterations.}
\centering \footnotesize
\begin{tabularx}{\textwidth}{l l|S[table-format=4.1] S[table-format=4.1] S[table-format=4.1] S[table-format=4.1] S[table-format=4.1] S[table-format=4.1] }
\toprule
\textbf{Target} & \textbf{Mode} & \text{Start EIF} & \text{Boot} & \text{Runner init} & \text{Checkout} & \text{Configure} & \text{Build}\\
\midrule
\multirow{5}{*}{Clang (18.1.3)}
 & H & 0.0\si{\second} & 0.1\si{\second} & 3.7\si{\second} & 145.0\si{\second} & 48.4\si{\second} & 2793.7\si{\second}\\
 & HS & 0.0\si{\second} & 0.1\si{\second} & 4.2\si{\second} & 148.0\si{\second} & 48.5\si{\second} & 3211.3\si{\second}\\
 & E & 71.0\si{\second} & 3.1\si{\second} & 4.1\si{\second} & 137.2\si{\second} & 36.7\si{\second} & 3270.9\si{\second}\\
 & ES & 46.4\si{\second} & 9.1\si{\second} & 4.1\si{\second} & 117.2\si{\second} & 35.6\si{\second} & 3784.1\si{\second}\\
 & ES+ & 46.4\si{\second} & 9.1\si{\second} & 5.5\si{\second} & 132.8\si{\second} & 115.1\si{\second} & 4748.5\si{\second}\\
\midrule
\multirow{5}{*}{Kernel (6.8.0, gcc)}
 & H & 0.0\si{\second} & 0.1\si{\second} & 3.5\si{\second} & 86.5\si{\second} & 5.3\si{\second} & 238.2\si{\second}\\
 & HS & 0.0\si{\second} & 0.1\si{\second} & 4.1\si{\second} & 86.1\si{\second} & 5.6\si{\second} & 272.6\si{\second}\\
 & E & 70.8\si{\second} & 3.1\si{\second} & 3.9\si{\second} & 69.5\si{\second} & 4.8\si{\second} & 246.5\si{\second}\\
 & ES & 46.4\si{\second} & 9.1\si{\second} & 3.9\si{\second} & 68.9\si{\second} & 5.0\si{\second} & 279.4\si{\second}\\
 & ES+ & 46.4\si{\second} & 9.1\si{\second} & 5.7\si{\second} & 91.0\si{\second} & 9.6\si{\second} & 533.8\si{\second}\\
\midrule
\multirow{5}{*}{Kernel (6.8.0, clang)}
 & H & 0.0\si{\second} & 0.1\si{\second} & 3.8\si{\second} & 80.1\si{\second} & 7.5\si{\second} & 410.2\si{\second}\\
 & HS & 0.0\si{\second} & 0.1\si{\second} & 4.1\si{\second} & 81.3\si{\second} & 7.4\si{\second} & 423.0\si{\second}\\
 & E & 71.1\si{\second} & 3.1\si{\second} & 4.2\si{\second} & 69.3\si{\second} & 5.5\si{\second} & 453.3\si{\second}\\
 & ES & 46.4\si{\second} & 9.1\si{\second} & 4.0\si{\second} & 68.6\si{\second} & 5.9\si{\second} & 447.3\si{\second}\\
 & ES+ & 46.3\si{\second} & 9.1\si{\second} & 5.6\si{\second} & 90.8\si{\second} & 11.6\si{\second} & 735.7\si{\second}\\
\midrule
\multirow{5}{*}{GProlog (1.6.0)}
 & H & 0.0\si{\second} & 0.1\si{\second} & 3.8\si{\second} & 1.3\si{\second} & 3.7\si{\second} & 12.8\si{\second}\\
 & HS & 0.0\si{\second} & 0.1\si{\second} & 4.2\si{\second} & 1.3\si{\second} & 3.8\si{\second} & 14.9\si{\second}\\
 & E & 43.5\si{\second} & 3.1\si{\second} & 4.0\si{\second} & 1.3\si{\second} & 2.9\si{\second} & 13.5\si{\second}\\
 & ES & 29.5\si{\second} & 8.1\si{\second} & 3.7\si{\second} & 1.4\si{\second} & 2.8\si{\second} & 14.3\si{\second}\\
 & ES+ & 29.5\si{\second} & 8.1\si{\second} & 5.5\si{\second} & 2.4\si{\second} & 14.3\si{\second} & 27.4\si{\second}\\
\midrule
\multirow{5}{*}{Hello (2.10)}
 & H & 0.0\si{\second} & 0.1\si{\second} & 3.6\si{\second} & 0.6\si{\second} & 14.5\si{\second} & 0.9\si{\second}\\
 & HS & 0.0\si{\second} & 0.1\si{\second} & 3.8\si{\second} & 0.6\si{\second} & 15.0\si{\second} & 1.1\si{\second}\\
 & E & 43.7\si{\second} & 3.1\si{\second} & 3.8\si{\second} & 0.6\si{\second} & 11.5\si{\second} & 0.9\si{\second}\\
 & ES & 29.7\si{\second} & 8.1\si{\second} & 4.3\si{\second} & 0.7\si{\second} & 11.0\si{\second} & 1.0\si{\second}\\
 & ES+ & 29.4\si{\second} & 8.1\si{\second} & 5.7\si{\second} & 1.4\si{\second} & 50.2\si{\second} & 4.5\si{\second}\\
\midrule
\multirow{5}{*}{IPXE (1.21.1)}
 & H & 0.0\si{\second} & 0.1\si{\second} & 3.9\si{\second} & 0.6\si{\second} & 0.0\si{\second} & n/a\\
 & HS & 0.0\si{\second} & 0.1\si{\second} & 3.7\si{\second} & 0.6\si{\second} & 0.0\si{\second} & 52.9\si{\second}\\
 & E & 43.5\si{\second} & 3.1\si{\second} & 4.0\si{\second} & 0.7\si{\second} & 0.0\si{\second} & 45.8\si{\second}\\
 & ES & 29.6\si{\second} & 8.1\si{\second} & 3.9\si{\second} & 0.7\si{\second} & 0.0\si{\second} & 48.6\si{\second}\\
 & ES+ & 29.5\si{\second} & 8.1\si{\second} & 5.5\si{\second} & 1.6\si{\second} & 0.0\si{\second} & 196.4\si{\second}\\
\midrule
\multirow{5}{*}{Scheme48 (1.9.3)}
 & H & 0.0\si{\second} & 0.1\si{\second} & 3.7\si{\second} & 0.6\si{\second} & 3.5\si{\second} & 14.8\si{\second}\\
 & HS & 0.0\si{\second} & 0.1\si{\second} & 4.0\si{\second} & 0.7\si{\second} & 3.7\si{\second} & 15.9\si{\second}\\
 & E & 43.4\si{\second} & 3.1\si{\second} & 3.7\si{\second} & 0.7\si{\second} & 2.8\si{\second} & 15.2\si{\second}\\
 & ES & 29.5\si{\second} & 8.1\si{\second} & 3.9\si{\second} & 0.7\si{\second} & 2.7\si{\second} & 16.0\si{\second}\\
 & ES+ & 29.4\si{\second} & 8.1\si{\second} & 5.9\si{\second} & 1.6\si{\second} & 12.7\si{\second} & 20.0\si{\second}\\
\midrule
\multirow{5}{*}{NeoVIM (0.11.0)}
 & H & 0.0\si{\second} & 0.1\si{\second} & 3.7\si{\second} & 0.9\si{\second} & 66.1\si{\second} & 255.9\si{\second}\\
 & HS & 0.1\si{\second} & 0.1\si{\second} & 3.7\si{\second} & 0.9\si{\second} & 75.9\si{\second} & 184.9\si{\second}\\
 & E & 43.5\si{\second} & 3.1\si{\second} & 3.8\si{\second} & 0.9\si{\second} & 65.3\si{\second} & 158.9\si{\second}\\
 & ES & 29.8\si{\second} & 8.1\si{\second} & 3.9\si{\second} & 0.9\si{\second} & 70.0\si{\second} & 167.3\si{\second}\\
 & ES+ & 29.4\si{\second} & 8.1\si{\second} & 5.7\si{\second} & 2.2\si{\second} & 130.7\si{\second} & 311.7\si{\second}\\
\midrule
\multirow{5}{*}{LibSodium (1.0.20)}
 & H & 0.0\si{\second} & 0.1\si{\second} & 3.9\si{\second} & 1.1\si{\second} & 13.2\si{\second} & 20.2\si{\second}\\
 & HS & 0.0\si{\second} & 0.1\si{\second} & 3.8\si{\second} & 1.3\si{\second} & 15.1\si{\second} & 23.4\si{\second}\\
 & E & 43.8\si{\second} & 3.1\si{\second} & 3.7\si{\second} & 0.8\si{\second} & 10.6\si{\second} & 19.9\si{\second}\\
 & ES & 29.5\si{\second} & 8.1\si{\second} & 4.0\si{\second} & 0.8\si{\second} & 11.5\si{\second} & 22.1\si{\second}\\
 & ES+ & 29.6\si{\second} & 8.1\si{\second} & 5.9\si{\second} & 1.7\si{\second} & 46.1\si{\second} & 94.9\si{\second}\\
\midrule
\multirow{5}{*}{TinyCC (0.9.28)}
 & H & 0.0\si{\second} & 0.1\si{\second} & 3.6\si{\second} & 0.7\si{\second} & 0.1\si{\second} & 4.5\si{\second}\\
 & HS & 0.0\si{\second} & 0.1\si{\second} & 4.5\si{\second} & 0.7\si{\second} & 0.1\si{\second} & 5.7\si{\second}\\
 & E & 43.5\si{\second} & 3.1\si{\second} & 3.9\si{\second} & 0.7\si{\second} & 0.1\si{\second} & 5.6\si{\second}\\
 & ES & 29.4\si{\second} & 8.1\si{\second} & 4.2\si{\second} & 0.7\si{\second} & 0.1\si{\second} & 5.8\si{\second}\\
 & ES+ & 29.5\si{\second} & 8.1\si{\second} & 5.6\si{\second} & 1.6\si{\second} & 0.4\si{\second} & 8.5\si{\second}\\
\midrule
\multirow{5}{*}{Verifier Client}
 & H & 0.0\si{\second} & 0.1\si{\second} & 3.7\si{\second} & 0.8\si{\second} & 4.4\si{\second} & 69.4\si{\second}\\
 & HS & 0.0\si{\second} & 0.1\si{\second} & 4.0\si{\second} & 0.6\si{\second} & 4.4\si{\second} & 68.5\si{\second}\\
 & E & 43.4\si{\second} & 3.1\si{\second} & 3.9\si{\second} & 0.7\si{\second} & 3.2\si{\second} & 70.5\si{\second}\\
 & ES & 29.5\si{\second} & 8.1\si{\second} & 4.1\si{\second} & 0.6\si{\second} & 2.7\si{\second} & 70.4\si{\second}\\
 & ES+ & 29.4\si{\second} & 8.1\si{\second} & 5.6\si{\second} & 1.3\si{\second} & 7.3\si{\second} & 110.9\si{\second}\\
\midrule
\multirow{5}{*}{XZ Utils (5.6.3)}
 & H & 0.0\si{\second} & 0.1\si{\second} & 3.7\si{\second} & 0.7\si{\second} & 9.7\si{\second} & 13.6\si{\second}\\
 & HS & 0.0\si{\second} & 0.1\si{\second} & 3.8\si{\second} & 0.7\si{\second} & 11.1\si{\second} & 16.1\si{\second}\\
 & E & 43.5\si{\second} & 3.1\si{\second} & 3.9\si{\second} & 0.6\si{\second} & 8.1\si{\second} & 14.1\si{\second}\\
 & ES & 29.4\si{\second} & 8.1\si{\second} & 3.8\si{\second} & 0.6\si{\second} & 8.2\si{\second} & 14.8\si{\second}\\
 & ES+ & 29.4\si{\second} & 8.1\si{\second} & 5.9\si{\second} & 1.5\si{\second} & 40.9\si{\second} & 69.0\si{\second}\\
\bottomrule
\end{tabularx}
\label{tab:full-table:main}
\end{table*}
\begin{table*}[h]
\caption{Build times for the C-based ``XZ Utils across all modes and job number combinations. All values in seconds and averaged over three iterations.}
\centering \footnotesize
\begin{tabularx}{\textwidth}{l l|S[table-format=4.1] S[table-format=4.1] S[table-format=4.1] S[table-format=4.1] S[table-format=4.1] S[table-format=4.1] }
\toprule
\textbf{Target} & \textbf{Mode} & \text{Start EIF} & \text{Boot} & \text{Runner init} & \text{Checkout} & \text{Configure} & \text{Build}\\
\midrule
\multirow{5}{*}{XZ Utils (5.6.3) (j1)}
 & H & 0.1\si{\second} & 0.1\si{\second} & 3.6\si{\second} & 0.6\si{\second} & 9.7\si{\second} & 36.1\si{\second}\\
 & HS & 0.0\si{\second} & 0.1\si{\second} & 4.0\si{\second} & 0.6\si{\second} & 11.1\si{\second} & 43.3\si{\second}\\
 & E & 43.4\si{\second} & 3.1\si{\second} & 3.8\si{\second} & 0.7\si{\second} & 8.1\si{\second} & 36.3\si{\second}\\
 & ES & 29.5\si{\second} & 8.1\si{\second} & 4.1\si{\second} & 0.7\si{\second} & 8.2\si{\second} & 38.7\si{\second}\\
 & ES+ & 29.4\si{\second} & 8.1\si{\second} & 6.0\si{\second} & 1.5\si{\second} & 40.6\si{\second} & 86.3\si{\second}\\
\midrule
\multirow{5}{*}{XZ Utils (5.6.3) (j2)}
 & H & 0.0\si{\second} & 0.1\si{\second} & 3.6\si{\second} & 0.7\si{\second} & 9.7\si{\second} & 19.4\si{\second}\\
 & HS & 0.0\si{\second} & 0.1\si{\second} & 4.0\si{\second} & 0.7\si{\second} & 11.2\si{\second} & 23.1\si{\second}\\
 & E & 43.5\si{\second} & 3.1\si{\second} & 3.6\si{\second} & 0.6\si{\second} & 8.1\si{\second} & 20.4\si{\second}\\
 & ES & 29.4\si{\second} & 8.1\si{\second} & 3.7\si{\second} & 0.6\si{\second} & 8.3\si{\second} & 21.5\si{\second}\\
 & ES+ & 29.4\si{\second} & 8.1\si{\second} & 6.0\si{\second} & 1.6\si{\second} & 41.1\si{\second} & 69.2\si{\second}\\
\midrule
\multirow{5}{*}{XZ Utils (5.6.3) (j3)}
 & H & 0.0\si{\second} & 0.1\si{\second} & 3.9\si{\second} & 0.7\si{\second} & 9.7\si{\second} & 15.7\si{\second}\\
 & HS & 0.0\si{\second} & 0.1\si{\second} & 3.9\si{\second} & 0.7\si{\second} & 11.1\si{\second} & 18.6\si{\second}\\
 & E & 43.5\si{\second} & 3.1\si{\second} & 4.0\si{\second} & 0.6\si{\second} & 8.1\si{\second} & 16.3\si{\second}\\
 & ES & 29.4\si{\second} & 8.1\si{\second} & 3.9\si{\second} & 0.6\si{\second} & 8.3\si{\second} & 17.3\si{\second}\\
 & ES+ & 29.4\si{\second} & 8.1\si{\second} & 5.5\si{\second} & 1.5\si{\second} & 40.7\si{\second} & 66.9\si{\second}\\
\midrule
\multirow{5}{*}{XZ Utils (5.6.3) (j4)}
 & H & 0.0\si{\second} & 0.1\si{\second} & 3.6\si{\second} & 0.7\si{\second} & 9.7\si{\second} & 13.6\si{\second}\\
 & HS & 0.0\si{\second} & 0.1\si{\second} & 4.1\si{\second} & 0.7\si{\second} & 11.2\si{\second} & 16.1\si{\second}\\
 & E & 43.5\si{\second} & 3.1\si{\second} & 4.2\si{\second} & 0.6\si{\second} & 8.2\si{\second} & 14.2\si{\second}\\
 & ES & 29.4\si{\second} & 8.1\si{\second} & 3.7\si{\second} & 0.7\si{\second} & 8.3\si{\second} & 14.9\si{\second}\\
 & ES+ & 29.4\si{\second} & 8.1\si{\second} & 5.5\si{\second} & 1.5\si{\second} & 40.7\si{\second} & 69.0\si{\second}\\
\midrule
\multirow{5}{*}{XZ Utils (5.6.3) (j5)}
 & H & 0.0\si{\second} & 0.1\si{\second} & 3.5\si{\second} & 0.7\si{\second} & 9.7\si{\second} & 13.7\si{\second}\\
 & HS & 0.0\si{\second} & 0.1\si{\second} & 3.8\si{\second} & 0.7\si{\second} & 11.2\si{\second} & 16.2\si{\second}\\
 & E & 43.4\si{\second} & 3.1\si{\second} & 3.8\si{\second} & 0.7\si{\second} & 8.1\si{\second} & 14.3\si{\second}\\
 & ES & 29.4\si{\second} & 8.1\si{\second} & 3.8\si{\second} & 0.7\si{\second} & 8.2\si{\second} & 14.8\si{\second}\\
 & ES+ & 29.5\si{\second} & 8.1\si{\second} & 5.7\si{\second} & 1.6\si{\second} & 41.0\si{\second} & 70.5\si{\second}\\
\midrule
\multirow{5}{*}{XZ Utils (5.6.3) (j6)}
 & H & 0.0\si{\second} & 0.1\si{\second} & 3.4\si{\second} & 0.7\si{\second} & 9.8\si{\second} & 13.8\si{\second}\\
 & HS & 0.0\si{\second} & 0.1\si{\second} & 3.9\si{\second} & 0.7\si{\second} & 11.2\si{\second} & 16.3\si{\second}\\
 & E & 43.8\si{\second} & 3.1\si{\second} & 3.9\si{\second} & 0.6\si{\second} & 8.2\si{\second} & 14.4\si{\second}\\
 & ES & 29.5\si{\second} & 8.1\si{\second} & 4.0\si{\second} & 0.6\si{\second} & 8.3\si{\second} & 15.2\si{\second}\\
 & ES+ & 29.8\si{\second} & 8.1\si{\second} & 5.9\si{\second} & 1.6\si{\second} & 41.3\si{\second} & 71.2\si{\second}\\
\midrule
\multirow{5}{*}{XZ Utils (5.6.3) (j7)}
 & H & 0.0\si{\second} & 0.1\si{\second} & 3.4\si{\second} & 0.7\si{\second} & 9.7\si{\second} & 13.8\si{\second}\\
 & HS & 0.0\si{\second} & 0.1\si{\second} & 4.1\si{\second} & 0.7\si{\second} & 11.1\si{\second} & 16.4\si{\second}\\
 & E & 43.6\si{\second} & 3.1\si{\second} & 3.9\si{\second} & 0.7\si{\second} & 8.1\si{\second} & 14.5\si{\second}\\
 & ES & 29.4\si{\second} & 8.1\si{\second} & 3.9\si{\second} & 0.7\si{\second} & 8.3\si{\second} & 15.2\si{\second}\\
 & ES+ & 29.5\si{\second} & 8.1\si{\second} & 5.6\si{\second} & 1.5\si{\second} & 40.4\si{\second} & 71.2\si{\second}\\
\midrule
\multirow{5}{*}{XZ Utils (5.6.3) (j8)}
 & H & 0.0\si{\second} & 0.1\si{\second} & 3.5\si{\second} & 0.7\si{\second} & 9.7\si{\second} & 13.7\si{\second}\\
 & HS & 0.0\si{\second} & 0.1\si{\second} & 3.9\si{\second} & 0.7\si{\second} & 11.2\si{\second} & 16.4\si{\second}\\
 & E & 43.5\si{\second} & 3.1\si{\second} & 3.7\si{\second} & 0.7\si{\second} & 8.1\si{\second} & 14.5\si{\second}\\
 & ES & 29.4\si{\second} & 8.1\si{\second} & 4.1\si{\second} & 0.6\si{\second} & 8.3\si{\second} & 15.2\si{\second}\\
 & ES+ & 29.4\si{\second} & 8.1\si{\second} & 5.6\si{\second} & 1.5\si{\second} & 41.1\si{\second} & 72.2\si{\second}\\
\bottomrule
\end{tabularx}
\label{tab:full-table:scalar-xz}
\end{table*}
\begin{table*}[h]
\caption{Build times for the Rust-based ``Verifier Client'' across all modes and job number combinations. All values in seconds and averaged over three iterations.}
\centering \footnotesize
\begin{tabularx}{\textwidth}{l l|S[table-format=4.1] S[table-format=4.1] S[table-format=4.1] S[table-format=4.1] S[table-format=4.1] S[table-format=4.1] }
\toprule
\textbf{Target} & \textbf{Mode} & \text{Start EIF} & \text{Boot} & \text{Runner init} & \text{Checkout} & \text{Configure} & \text{Build}\\
\midrule
\multirow{5}{*}{Verifier Client (j1)}
 & H & 0.0\si{\second} & 0.1\si{\second} & 3.6\si{\second} & 0.6\si{\second} & 4.3\si{\second} & 186.5\si{\second}\\
 & HS & 0.0\si{\second} & 0.1\si{\second} & 3.9\si{\second} & 0.6\si{\second} & 4.6\si{\second} & 185.1\si{\second}\\
 & E & 43.4\si{\second} & 3.1\si{\second} & 3.9\si{\second} & 0.6\si{\second} & 3.5\si{\second} & 181.2\si{\second}\\
 & ES & 29.4\si{\second} & 8.1\si{\second} & 3.9\si{\second} & 0.6\si{\second} & 2.7\si{\second} & 181.2\si{\second}\\
 & ES+ & 29.5\si{\second} & 8.1\si{\second} & 5.8\si{\second} & 1.4\si{\second} & 7.3\si{\second} & 230.4\si{\second}\\
\midrule
\multirow{5}{*}{Verifier Client (j2)}
 & H & 0.0\si{\second} & 0.1\si{\second} & 3.6\si{\second} & 0.6\si{\second} & 4.4\si{\second} & 100.0\si{\second}\\
 & HS & 0.0\si{\second} & 0.1\si{\second} & 4.1\si{\second} & 0.6\si{\second} & 4.2\si{\second} & 99.1\si{\second}\\
 & E & 43.4\si{\second} & 3.1\si{\second} & 3.9\si{\second} & 0.6\si{\second} & 3.1\si{\second} & 97.8\si{\second}\\
 & ES & 29.4\si{\second} & 8.1\si{\second} & 3.8\si{\second} & 0.6\si{\second} & 2.8\si{\second} & 98.3\si{\second}\\
 & ES+ & 29.4\si{\second} & 8.1\si{\second} & 5.9\si{\second} & 1.4\si{\second} & 7.3\si{\second} & 138.8\si{\second}\\
\midrule
\multirow{5}{*}{Verifier Client (j3)}
 & H & 0.0\si{\second} & 0.1\si{\second} & 3.4\si{\second} & 0.6\si{\second} & 4.3\si{\second} & 80.2\si{\second}\\
 & HS & 0.0\si{\second} & 0.1\si{\second} & 3.9\si{\second} & 0.6\si{\second} & 4.2\si{\second} & 79.4\si{\second}\\
 & E & 43.4\si{\second} & 3.1\si{\second} & 7.1\si{\second} & 0.6\si{\second} & 3.2\si{\second} & 80.1\si{\second}\\
 & ES & 29.4\si{\second} & 8.1\si{\second} & 4.0\si{\second} & 0.6\si{\second} & 2.7\si{\second} & 80.6\si{\second}\\
 & ES+ & 29.4\si{\second} & 8.1\si{\second} & 5.9\si{\second} & 1.3\si{\second} & 7.3\si{\second} & 114.0\si{\second}\\
\midrule
\multirow{5}{*}{Verifier Client (j4)}
 & H & 0.0\si{\second} & 0.1\si{\second} & 3.7\si{\second} & 0.7\si{\second} & 4.3\si{\second} & 69.2\si{\second}\\
 & HS & 0.0\si{\second} & 0.1\si{\second} & 4.1\si{\second} & 0.6\si{\second} & 4.7\si{\second} & 68.3\si{\second}\\
 & E & 43.4\si{\second} & 3.1\si{\second} & 3.6\si{\second} & 0.6\si{\second} & 3.2\si{\second} & 70.2\si{\second}\\
 & ES & 29.4\si{\second} & 8.1\si{\second} & 3.8\si{\second} & 0.7\si{\second} & 2.7\si{\second} & 70.4\si{\second}\\
 & ES+ & 29.4\si{\second} & 8.1\si{\second} & 5.6\si{\second} & 1.4\si{\second} & 7.5\si{\second} & 110.6\si{\second}\\
\midrule
\multirow{5}{*}{Verifier Client (j5)}
 & H & 0.0\si{\second} & 0.1\si{\second} & 3.6\si{\second} & 0.6\si{\second} & 4.2\si{\second} & 69.5\si{\second}\\
 & HS & 0.0\si{\second} & 0.1\si{\second} & 3.8\si{\second} & 0.6\si{\second} & 4.4\si{\second} & 68.7\si{\second}\\
 & E & 43.5\si{\second} & 3.1\si{\second} & 3.8\si{\second} & 0.7\si{\second} & 3.4\si{\second} & 70.7\si{\second}\\
 & ES & 29.5\si{\second} & 8.1\si{\second} & 4.1\si{\second} & 0.6\si{\second} & 2.9\si{\second} & 71.7\si{\second}\\
 & ES+ & 29.7\si{\second} & 8.1\si{\second} & 5.6\si{\second} & 1.3\si{\second} & 7.5\si{\second} & 112.8\si{\second}\\
\midrule
\multirow{5}{*}{Verifier Client (j6)}
 & H & 0.0\si{\second} & 0.1\si{\second} & 3.6\si{\second} & 1.0\si{\second} & 4.3\si{\second} & 70.8\si{\second}\\
 & HS & 0.0\si{\second} & 0.1\si{\second} & 3.8\si{\second} & 0.6\si{\second} & 4.4\si{\second} & 69.7\si{\second}\\
 & E & 43.8\si{\second} & 3.1\si{\second} & 3.8\si{\second} & 0.6\si{\second} & 4.2\si{\second} & 72.2\si{\second}\\
 & ES & 29.4\si{\second} & 8.1\si{\second} & 3.9\si{\second} & 0.7\si{\second} & 2.7\si{\second} & 72.1\si{\second}\\
 & ES+ & 29.7\si{\second} & 8.1\si{\second} & 5.8\si{\second} & 1.3\si{\second} & 7.3\si{\second} & 115.3\si{\second}\\
\midrule
\multirow{5}{*}{Verifier Client (j7)}
 & H & 0.0\si{\second} & 0.1\si{\second} & 3.7\si{\second} & 0.6\si{\second} & 4.2\si{\second} & 70.4\si{\second}\\
 & HS & 0.0\si{\second} & 0.1\si{\second} & 4.2\si{\second} & 0.6\si{\second} & 4.6\si{\second} & 69.3\si{\second}\\
 & E & 43.4\si{\second} & 3.1\si{\second} & 3.8\si{\second} & 0.7\si{\second} & 3.8\si{\second} & 72.3\si{\second}\\
 & ES & 29.4\si{\second} & 8.1\si{\second} & 3.6\si{\second} & 0.6\si{\second} & 3.2\si{\second} & 72.1\si{\second}\\
 & ES+ & 29.5\si{\second} & 8.1\si{\second} & 5.7\si{\second} & 1.3\si{\second} & 7.5\si{\second} & 117.0\si{\second}\\
\midrule
\multirow{5}{*}{Verifier Client (j8)}
 & H & 0.0\si{\second} & 0.1\si{\second} & 3.5\si{\second} & 0.6\si{\second} & 4.2\si{\second} & 71.3\si{\second}\\
 & HS & 0.0\si{\second} & 0.1\si{\second} & 3.9\si{\second} & 0.6\si{\second} & 4.2\si{\second} & 70.3\si{\second}\\
 & E & 43.4\si{\second} & 3.1\si{\second} & 3.8\si{\second} & 0.6\si{\second} & 3.1\si{\second} & 73.1\si{\second}\\
 & ES & 29.4\si{\second} & 8.1\si{\second} & 4.0\si{\second} & 0.6\si{\second} & 3.2\si{\second} & 73.3\si{\second}\\
 & ES+ & 29.4\si{\second} & 8.1\si{\second} & 5.9\si{\second} & 1.4\si{\second} & 7.4\si{\second} & 120.0\si{\second}\\
\bottomrule
\end{tabularx}
\label{tab:full-table:scalar-vc}
\end{table*}

\fi 

\end{document}